\documentclass[fleqn,10pt]{wlscirep}
\usepackage[utf8]{inputenc}
\usepackage[T1]{fontenc}
\usepackage{graphicx}
\usepackage{color}
\usepackage{dcolumn}
\usepackage{latexsym}
\usepackage[normalem]{ulem}
\usepackage{hyperref,amssymb}
\usepackage{url}
\usepackage{pifont}
\usepackage{color,soul}
\usepackage{graphicx}
\usepackage{verbatim}
\usepackage{multirow}
\usepackage{amsmath}
\newcommand{\beq}{\begin{eqnarray}}
\newcommand{\eeq}{\end{eqnarray}}
\usepackage{mathrsfs}
\usepackage{float,soul}
\usepackage{mathtools}
\usepackage{slashed}	
\usepackage{graphicx}   
\usepackage{epstopdf}
\usepackage{tabularx}
\usepackage{subfigure}  
\usepackage{hyperref}   
\usepackage{bbold}
\usepackage{wasysym}
\usepackage{feynmp}
\usepackage{hyperref}
\hypersetup{colorlinks,
}

\usepackage{dcolumn}
\usepackage{latexsym}
\usepackage{tikz}
\DeclareGraphicsExtensions{.pdf,.png,.jpg,.eps}

\newcommand{\angstrom}{\textup{\AA}}
\usepackage{siunitx}
\usepackage{soul}
\usepackage{relsize}
\soulregister{\cite}7
\soulregister{\ref}7
\soulregister{\eqref}7
\usepackage{amsmath}

\usepackage{bm}
\usepackage{booktabs}

\usepackage{xcolor}
\DeclareSIUnit\angstrom{\text{\AA}}

\title{Three-stage melting of a macroscopic continuous spacetime crystal}

\author[1,2,$\dagger$]{Guoqing Liu}
\author[1,3,$\dagger$]{Jimin Bai}
\author[1,3,*]{Matteo Baggioli}
\author[1,2,*]{Jie Zhang}
\affil[1]{School of Physics and Astronomy, Shanghai Jiao Tong University, Shanghai 200240, China}
\affil[2]{Institute of Natural Sciences,Shanghai Jiao Tong University, Shanghai 200240, China}
\affil[3]{Wilczek Quantum Center, Shanghai Jiao Tong University, Shanghai 200240, China}

\affil[$\dagger$]{These authors contributed equally to this work}
\affil[*]{Corresponding authors: \color{blue}b.matteo@sjtu.edu.cn\color{black}, 
 \color{blue}jiezhang2012@sjtu.edu.cn\color{black}}

\begin{abstract}
\textbf{A spacetime crystal is a phase of matter that spontaneously develops periodic order in both space and time. Spacetime crystals have been experimentally observed in microscopic quantum many-body systems and, very recently, in a mesoscopic nematic liquid crystal. However, the melting process of a spacetime crystal and its underlying physical mechanisms have not yet been experimentally reported. Here, we present a direct observation of a classical continuous spacetime crystal melting in a table-top experiment with macroscopic active granular disks in $2+1$ spacetime dimensions. The spacetime crystal is characterized by the spontaneous formation of a coherent, rigid-body rotation of a 2D triangular lattice that persists for almost a day and remains remarkably robust to noise. By tuning the disk packing fraction, we observe a complex three-stage melting process involving a spatially hexatic phase and multiple coexistence regions. Importantly, we show that spatial and temporal crystalline orders melt separately through distinct mechanisms: spatial order is destroyed by the proliferation of topological defects, while temporal order is lost through the decay of directional persistence caused by the progressive weakening of many-body interactions. Our results demonstrate that the spontaneous breaking of spatial and temporal translational symmetries can be decoupled, leading to the emergence of exotic out-of-equilibrium classical phases of matter.}
\end{abstract}
\begin{document}

\flushbottom
\maketitle

\thispagestyle{empty}

Crystallization and melting are classic manifestations of the spontaneous emergence and loss of order, observed across both natural systems and technological processes. At their core, these processes reflect the breaking and restoration of continuous spatial translational symmetry: a fluid crystallizes when continuous symmetry gives way to a discrete periodic lattice, and it melts when that ordered structure is lost \cite{kittel2018introduction, chaikin1995principles}. Remarkably, even systems as simple as hard-sphere fluids can crystallize purely through entropic effects \cite{frenkel2015order}, as first demonstrated in seminal theoretical and experimental work \cite{wood1957preliminary, alder1957phase}. In two spatial dimensions, this phenomenon is even richer, with melting proceeding through a two-stage process mediated by the proliferation of topological defects and the appearance of an intermediate hexatic phase \cite{2dmelt-RevModPhys.60.161, JM-Kosterlitz_1973, Nelson-PhysRevB.19.2457, Young-PhysRevB.19.1855, nelson2002defects}.

The profound analogy between space and time in fundamental physics has motivated the search for analogous symmetry-breaking patterns in the temporal domain. This idea was crystallized by Frank Wilczek’s proposal of a ``time crystal,” a phase of matter that spontaneously breaks continuous time-translational symmetry \cite{FW-PhysRevLett.109.160401}. Although subsequent no-go theorems \cite{nozieres2013time, PhysRevLett.111.070402, PhysRevLett.114.251603} ruled out such behavior in isolated equilibrium systems, they also revealed that time-crystalline order can emerge in driven systems, sparking intense theoretical debate and experimental exploration \cite{RevModPhys.95.031001,Sacha_2018,khemani2019brief}. These efforts culminated in the first observations of discrete and continuous time crystals \cite{zhang2017observation, kongkhambut2022observation}, followed by demonstrations across a wide range of quantum platforms \cite{choi2017observation, PhysRevLett.120.180603, PhysRevLett.121.185301, PhysRevLett.120.215301, O’Sullivan_2020, kyprianidis2021observation, doi:10.1126/science.abk0603, mi2022time, chen2023realization, wu2024dissipative, greilich2024robust, doi:10.1126/science.adn7087}.

More recently, time-crystalline order has been shown not to be restricted to quantum systems. Both continuous and discrete time crystals have been realized in classical settings, including systems of liquid-crystal molecules \cite{zhao2025space, zhao2025emergent}, acoustically levitated particles with nonreciprocal interactions \cite{zjzk-t81n}, Hamiltonian systems \cite{PhysRevLett.109.160402,Dai_2020} and so on \cite{PhysRevLett.123.124301,Yao2020,liu2023photonic}. At the same time, the existence of spacetime crystals that simultaneously break translational symmetry in both space and time has been established \cite{PhysRevLett.121.185301,PhysRevLett.109.163001,PhysRevLett.126.057201,zhao2025space}, giving rise to a rich and intricate hierarchy of symmetry-breaking patterns \cite{PhysRevLett.120.096401}.

Despite this rapid progress, a fundamental question remains largely unexplored: how does a time crystal lose its order? Unlike the extensively studied melting of ordinary spatial crystals, the melting of a spacetime crystal—and, crucially, its direct experimental observation—has so far remained elusive, confined largely to a few idealized theoretical toy models \cite{PhysRevB.105.L100303,PhysRevLett.131.221601}. In particular, it remains an open question whether spatial and temporal order melt through distinct physical mechanisms, and how temporal rigidity intertwines with topological melting scenarios in two spatial dimensions.

\subsection*{Experimental observation of a macroscopic spacetime crystal}

We report the experimental observation of a classical, macroscopic spacetime-crystalline phase in a quasi-two-dimensional (2D) system of active disk-shaped granular particles, referred to as \textit{granular disks} for simplicity.

The experimental setup, shown in Fig.~\ref{fig1}A, consists of a single layer of granular disks with disk-shaped ratchet caps (pitch diameter $D=8.8$ mm) supported with six alternatively inclined legs, placed on an aluminum plate and confined within a circular region of $\approx 50$ cm in diameter. The legs, with a height of $2.6$ mm, are alternatively deviated from the mid-axis plane by $\pm 60^\circ$. A small marker is placed on each particle's surface to indicate its orientation and enable precise tracking of its rotation. The granular disks are driven by a vertical sinusoidal vibration at $100$ Hz. Their motion is quasi-2D on the $(X,Y)$ plane and is tracked in real time using a Basler CCD camera at $40$ frames per second.

The active force on each granular disk is generated by collisions between the particle’s tilted legs and the vibrating bottom plane. The unique design of the legs and the presence of noise cause the particle’s central axis to slightly tilt away from the gravitational direction during vibration. This typically results in a single leg being instantaneously propelled upon contact with the bottom surface. Such interactions generate a horizontal force that depends on both the collision velocity and the angle of contact between the leg and the bottom surface. The tilted legs lead to minimal time correlation in the contact angle, resulting in a random orientation of the driving force.

Each particle independently absorbs energy from the vertical driving and dissipates it through inelastic collisions with neighboring particles and with the vibrating plate. The translational dynamics of an isolated particle, shown in Fig.~\ref{fig1}B, resemble Brownian motion, characterized by the absence of persistent directional movement. Theoretically, a single particle shall exhibit zero mean spin angular momentum over time, due to the symmetry in the inclination of its six legs. In contrast, a single disk exhibits a finite spin angular momentum, resulting from imperfections in particle manufacturing, as illustrated in Fig.~\ref{fig1}C. 

At a high disk packing fraction $\varphi = 0.835$, we observe the emergence of a synchronized rigid-body rotation, in which all disks rotate coherently in a periodic fashion for durations of up to one day. The packing fraction $\varphi$ is defined as the ratio between the total area occupied by the particles and the system area.
The trajectories of three representative disks are shown as blue-to-red arrows in Fig.~\ref{fig1}D, while videos of this emergent periodic motion are provided in the Supplementary Materials (SM). This spontaneous motion is observed universally and does not depend on particle type or chirality see Supplementary Movies~\ref{vid:835},~\ref{vid:MRS},~\ref{vid:CS} and~\ref{vid:RC}). 
Although individual disks’ spin angular momenta are mostly random and their average over different particles is typically negligibly small, we observe a consistent bias toward counterclockwise rotation, as illustrated in Fig.~\ref{fig1}D. As shown in the SM, this bias is neither dictated by particle shape nor by the arrangement of the ratchets, and it persists even for disk-cap particles of no chirality. This bias is most likely due to minor imperfections in the experimental setup that are amplified by many-body interactions at high packing fractions.

By tracking the individual particle coordinates $X_i(t)$ and $Y_i(t)$, we find that each disk undergoes periodic sinusoidal oscillations with a characteristic period $T \approx 4.70$ hours, as explicitly shown for two disks in Fig.~\ref{fig1}E. This period is roughly six orders of magnitude larger than that of the external vertical vibrations. This indicates that the latter merely inject kinetic energy at the particle level and are entirely unrelated to the onset of the time-crystalline phase. As we will show below, they in fact act against it when the disk packing fraction is reduced. Moreover, the external periodic vibrations act solely along the $z$ direction (see Fig.~\ref{fig1}A) and do not impose any periodic forcing in the $(X,Y)$ plane. The observed temporal ordering therefore emerges spontaneously in the transverse dynamics and corresponds to a continuous, rather than discrete, breaking of time-translation symmetry. We have repeated the experiment multiple times under identical conditions and verified that the emergent period is robust across independent realizations, consistently lying within the range of $4.7$–$5.5$ hours.

The corresponding Fourier power spectrum, shown in Fig.~\ref{fig1}F, exhibits a pronounced peak at $f \approx 0.55\times 10^{-4}$ Hz, confirming the highly periodic nature of the motion. Moreover, the time auto-correlation function of the normalized projection of the periodic motion $\tilde{y}$, $G(t)\equiv \langle \tilde{y}(t)\tilde{y}(0)\rangle-\langle \tilde{y}(t)\rangle \langle \tilde{y}(0)\rangle$ where the average refers to particles whose distance to the origin is between $8.8$ cm and $20.6$ cm to avoid the center and the boundary regions, reveals long-range temporal order, characterized by a slow power-law decay \color{black} $|G(t)| \sim t^{-0.09\pm0.05}$. 

To establish the time-crystalline nature of this state and its underlying rigidity, we demonstrate its robustness against strong external perturbations by injecting intense noise using a high-power acoustic woofer (see the Supplementary Movie~\ref{vid:noise}). In addition, measurements on seven nominally identical systems, all driven simultaneously, demonstrate that the emergence of the time-crystalline state is highly nontrivial and spontaneous in nature. Each system has a diameter of $15$ cm and is arranged uniformly within a larger circle of diameter $50$ cm, as shown in the SM. Notably, two of the systems do not develop the time-crystalline state during the $15$-hour experimental run, indicating that their onset times exceed the finite observation window. Among the systems that do exhibit a time-crystalline state, both the onset time relative to the start of the external driving and the phase of the resulting periodic motion are random. See Fig.~\ref{SM1} in the SM and related discussion for more details. The combination of noise robustness and random phase selection, documented in detail in the SM, together with supporting videos (see Supplementary Movies~\ref{vid:noise} and~\ref{vid:random_phase}), confirms the spontaneous breaking of continuous-time translational symmetry and validates the identification of this state as a continuous time crystal. This interpretation is further validated in the SM by the direct observation of an emergent Goldstone mode corresponding to gapless phase fluctuations. As with the large $50$-cm system, at $\varphi = 0.835$, the chirality of the rigid-body motion in these small $15$-cm systems is not selected spontaneously but rather arises from external biases. Consequently, chiral symmetry is not spontaneously broken.
 
Self-spinning crystals have been observed in a wide range of active systems, from swimming starfish embryos \cite{tan2022odd} to magnetic colloids \cite{bililign2022motile}. This behavior has been interpreted within the framework of odd elasticity \cite{scheibner2020odd, fruchart2023odd}, which arises from nonreciprocal transverse interactions between neighboring particles, similar to those observed in acoustically levitated systems \cite{zjzk-t81n}. In 2D crystals, such interactions largely cancel in the bulk but might remain finite at the boundary, possibly resulting in a net surface torque \cite{huang2025anomalous}. However, observations in the T- and S-coexistence regimes ($0.687<\varphi <0.734$, as will be introduced in Fig.~\ref{fig2}) contradict the above interpretation: the global rotation direction is random across repeated runs, indicating that nonreciprocal transverse interactions due to ratchet, if they exist, are not the primary origin of the rotation. This also excludes the small experimental biases observed at the highest packing fraction, $\varphi = 0.835$, deep in the spacetime crystal phase, as the underlying driving mechanism.

Instead, the global rotation of the spacetime crystal originates from emergent flocking of the active granular disks confined within the circular domain. This interpretation is consistent with recent experiments demonstrating flocking in active, nonpolar granular Brownian vibrators \cite{chen2024anomalous}. The geometry of the confinement is essential. A circular boundary promotes coherent collective motion and enables the global rotation of the crystal, whereas a flower-shaped boundary suppresses this effect by disrupting large-scale alignment \cite{chen2024anomalous}. Consistently, global rotation has been recovered in circular domains for systems composed of nonchiral disk-cap granular particles, similar to those studied in \cite{chen2024anomalous}, though smaller in size (see SM). The relevant timescale is equally important. At short times, inelastic collisions are infrequent, interactions remain weak, and particle motion is essentially random \cite{chen2024anomalous, doi:10.1073/pnas.2510873123}. At longer times, the accumulation of collisions generates an effective mutual attraction that progressively overcomes single-particle activity, leading to collective alignment and flocking \cite{chen2024anomalous}. Thus, under circular confinement, the competition between packing-fraction–controlled activity and collision-induced interactions drives a transition to global rotation, analogous to the coordinated motion observed in schools of fish or flocks of birds \cite{TONER2005170}.

At this highest packing fraction of $\varphi=0.835$, the particles self-organize into a periodic triangular lattice, clearly visible in the snapshot shown in Fig.~\ref{fig1}G. After subtracting the rigid-body rotation described above, the residual dynamics consist of collective vibrations around well-defined equilibrium positions across a range of spatial and temporal scales. These fluctuations correspond to collective transverse and longitudinal phonon modes, which are directly reflected in the intensity of the dynamical structure factors shown in Fig.~\ref{fig1}H.

Besides the dynamics, the structures also exhibit prominent features of a 2D spatially ordered state. The translational correlation function (see SM) exhibits a power-law decay, $G_T(r) \sim r^{-0.0545\pm 0.0020}$, while the bond-orientational correlation function (or hexatic correlation function) $G_6(r)$ displays long-range order in both space and time, as will be discussed in detail later in Fig.~\ref{fig4}. Together, these structural and dynamical signatures demonstrate that the system resides in a spatially ordered phase with translational quasi–long-range order, as expected for a 2D crystal. 

When combined with the temporal ordering discussed above, these results establish that our experimental system realizes a classical continuous spacetime crystal. Notably, owing to its macroscopic size, this phase is directly visible to the naked eye and remains stable for nearly a day, with the duration limited by the experimental apparatus.

\subsection*{Three-stage melting}
We investigate the melting of the spacetime crystal phase by systematically varying the packing fraction of the granular disks, $\varphi$. To independently quantify spatial and temporal order, we introduce the \textit{spatial crystalline fraction}, defined as the normalized intensity of the Bragg peaks at the high-symmetry points of the triangular lattice, and the \textit{time-crystalline fraction}, defined as the normalized intensity of the time-crystalline peak in the frequency power spectrum (see Methods for details). By construction, a crystalline fraction equal to $1$ corresponds to a perfect crystal, while a value of $0$ indicates the complete absence of crystalline order.

By decreasing $\varphi$ from the highest packing fraction shown in Fig.~\ref{fig1}, we identify four distinct phases as presented in Fig.~\ref{fig2}A.
(i) \textit{Spacetime crystal}: For $\varphi>\varphi_3 \approx 0.734$ (brick-red region), the system forms a spacetime crystalline phase in which both crystalline fractions are nearly unity and essentially independent of $\varphi$.
(ii) \textit{T-coexistence region}: In the interval $\varphi_2<\varphi<\varphi_3$ (pink region), with $\varphi_2 \approx 0.709$, the time-crystalline fraction drops rapidly and vanishes around $\varphi_2$, while the spatial crystalline fraction decreases more gradually and remains finite at $\varphi_2$.
(iii) \textit{S-coexistence region}: For $\varphi_1<\varphi<\varphi_2$ (mint region), with $\varphi_1 \approx 0.687$, the spatial crystalline fraction continues to decrease and eventually vanishes at $\varphi_1$.
(iv) \textit{Fluid}: For $\varphi<\varphi_1$ (green color), both spatial and temporal crystalline fractions vanish identically.

In Fig.~\ref{fig2}B, we present the real-time dynamics of the system by tracking the displacement of a selected group of particles over approximately $40$ minutes. Extended data and videos are provided in the SM, Supplementary Movies~\ref{vid:662},~\ref{vid:699},~\ref{vid:719} and SM Section~\ref{SM4}. The results must be compared with the periodic time-crystalline motion shown in Fig.~\ref{fig1}D, which persists down to $\varphi=\varphi_3$. As the packing fraction decreases further, the circular trajectories become increasingly noisy. Initially, only some trajectories lose their coherent motion, suggesting a region of phase coexistence (T-coexistence), which we will confirm through a more detailed analysis below. When $\varphi$ drops below $\varphi_2$, the random motion begins to influence the entire system and all particle trajectories. Trajectories are no longer purely tangential, but develop a persistent component along the radial direction. Ultimately, below $\varphi_1$, the motion of the particles becomes completely random and Brownian, dominated by the injected kinetic energy, similar to what is seen in a dilute fluid.

This initial characterization already indicates that the time-crystalline phase is a genuine many-body effect, arising from strong interparticle interactions that overcome the random kinetic energy injected by the external vertical vibrations. These interactions, together with the confining geometry, generate long-range temporal synchronization at sufficiently large time scales, ultimately stabilizing a robust time-crystalline phase by converting individual particle motions into collective angular momentum. This mechanism is further highlighted by the particle mean-square displacement (MSD) shown in Fig.~\ref{fig2}C. At low packing fractions, the MSD exhibits a standard long-time diffusive behavior, $\mathrm{MSD}\sim t$. In contrast, within the time-crystalline phase at large $\varphi$, the scaling changes dramatically to a ballistic form, $\mathrm{MSD}\sim t^{2}$. As derived analytically in the SM, this behavior provides direct evidence of the rigid-body periodic motion underlying this phase.

The loss of spatial order is examined further in Fig.~\ref{fig2}D, which shows the intensity of the 2D static structure factor $S(q_x,q_y)$ of the four packing fractions shown in Fig.~\ref{fig2}C. At the highest packing fraction, sharp Bragg peaks appear, confirming the presence of a triangular lattice with quasi–long-range translational order. In the interval $\varphi_2<\varphi<\varphi_3$, the Bragg peaks broaden and gradually diminish. Nevertheless, the sixfold rotational symmetry remains intact, signaling the emergence of a hexatic phase. As the packing fraction decreases further, this hexatic order is gradually destroyed through a coexistence region and disappears below $\varphi=\varphi_1$. As discussed in detail below, this hexatic-to-fluid transition involves a spatial coexistence regime and deviates from the standard continuous KTHNY scenario \cite{2dmelt-RevModPhys.60.161, JM-Kosterlitz_1973, Nelson-PhysRevB.19.2457, Young-PhysRevB.19.1855, nelson2002defects}.

\subsection*{Melting of time-crystalline order}
To characterize the system's dynamical behavior, we introduce a dimensionless metric that measures how particles move collectively in space over time. Specifically, we use the normalized non-affine displacement measure $\log_{10}\!\left(D_{\min}^2/D^2\right)$ \cite{PhysRevE.57.7192}, which captures local deviations from affine motion over a time lag $\Delta t$ (see Methods). The probability distribution functions (PDF) of the particle-level $\log_{10}\!\left(D_{\min}^2/D^2\right)$ are presented in Fig.~\ref{fig3}A for four different phases as depicted in Fig.~\ref{fig2}. In simple terms, this quantity probes the degree of dynamical coherence of a particle relative to its neighbors. Regions with small $\log_{10}\!\left(D_{\min}^2/D^2\right)$ correspond to highly coherent motion, as illustrated by the blue box in Fig.~\ref{fig3}B, where the collective particle dynamics are clearly visible. In contrast, regions with large $\log_{10}\!\left(D_{\min}^2/D^2\right)$, such as the red box in the same panel, exhibit fully disordered particle motion without temporal synchronization.

In the spacetime crystalline phase, the PDF of $\log_{10}\!\left(D_{\min}^2/D^2\right)$ displays a single peak at small values, reflecting uniformly coherent dynamics. In contrast, within the T-coexistence regime (pink region in the phase diagram of Fig.~\ref{fig2}A), the PDF becomes bimodal, with two well-separated peaks. This bimodality provides clear evidence for dynamical coexistence between a time-crystalline phase and a time-disordered fluid phase. These two dynamical states are directly visible in the snapshot shown in Fig.~\ref{fig3}B, where a sharp interface separates a coherent time-crystalline region (blue) from a time-disordered region (red). A quantitative analysis of the time-crystalline clusters within the coexistence regime shows that they vanish sharply at $\varphi_2$, confirming that for $\varphi<\varphi_2$ the system is dynamically disordered and fluid-like from a temporal-order perspective, as shown in Fig.~\ref{fig3}C.

To further corroborate this picture, we introduce an independent dynamical probe: the degree of directional persistence, which quantifies long-time correlations in the direction of particle velocities in the co-rotating frame (see Methods). In the spacetime crystalline phase, the directional persistence is maximal and independent of the packing fraction $\varphi$, consistent with coherent rigid-body rotation, as shown in Fig.~\ref{fig3}D. Within the T-coexistence region, directional persistence progressively decreases as dynamically disordered clusters emerge and proliferate. At $\varphi=\varphi_2$, these disordered regions span the entire system, and the directional persistence drops to zero, except for small intermittent fluctuations. For $\varphi<\varphi_1$, directional persistence is completely lost, and the time dynamics becomes completely random, as expected for a dilute fluid.

\subsection*{Melting of 2D spatial order via the proliferation of topological defects}
After elucidating the particle-level dynamics responsible for the melting of the time-crystalline order, we turn into the the melting of the spatial counterpart by analyzing structural properties and the corresponding defects. 

First, in panels A and B of Fig.~\ref{fig4}, we analyze the hexatic order parameter $\psi_6$ (see Methods) in the S-coexistence region, highlighted in mint in the phase diagram of Fig.~\ref{fig2}A. In this phase, the time-crystalline order is completely lost and the dynamics are fully disordered and fluid-like. Despite this, the spatial structure reveals a clear coexistence of hexatic regions with large $|\psi_6|$ values (red) and fluid-like clusters characterized by very small $|\psi_6|$ (blue), as shown in Fig.~\ref{fig4}A. Here, $|\psi_6|$ refers to the modulus of $\psi_6$. This spatial coexistence is further supported by the PDF of the complex hexatic order parameter $\psi_6$, which displays a characteristic bimodal distribution, as shown in Fig.~\ref{fig4}B.

In Fig.~\ref{fig4}C, we quantify the loss of orientational order by monitoring the percolation of $\psi_6$ clusters. In the crystalline and hexatic phases (brick-red and pink regions) above $\varphi_2$, the structural percolation parameter is maximal and essentially independent of the packing fraction. Entering the S-coexistence region, this parameter drops sharply, signaling the growth of fluid-like clusters with weak or absent orientational order. Finally, below $\varphi_1$ (green region), orientational order is completely lost and the system behaves as a fluid in both spatial and temporal dimensions.

To further characterize the extent of spatial structural order, we analyze the space and time correlation functions of the hexatic order parameter, $G_6(r)$ and $G_6(t)$ (see Methods), shown respectively in panels D and E of Fig.~\ref{fig4}. For completeness, the SM also report the behavior of the translational pair correlation function $g(r)$, which is fully consistent with the results discussed in the main text.
In the spatial crystalline phase (brick-red region), both the spatial and temporal correlations of the hexatic order parameter show no decay, confirming the presence of a true long-range orientational order, as expected for a 2D crystal. In contrast, within the T-coexistence region, both correlations decay algebraically, with exponents compatible with quasi-long-range hexatic order. From a structural perspective, the T-coexistence region therefore corresponds to a hexatic phase.
Below $\varphi_2$, both correlation functions show a clear exponential decay, with a correlation length that progressively decreases with lowering the packing fraction. This behavior signals the loss of orientational order and the system's melting into a fluid phase.

Finally, in Fig.~\ref{fig4}F, we present a detailed analysis of structural defects and their proliferation. Using a Voronoi decomposition of the configurations at each packing fraction (see Methods and SM), we identify isolated disclinations (5- or 7-fold coordinated particles), bound dislocations (5–7 pairs), and extended clusters of topological defects. Aside from a small number of randomly distributed defects, the spacetime crystalline phase at large $\varphi$ shows no evidence of free defects.

In contrast, in the hexatic phase (pink region), we observe a clear proliferation of free dislocations, consistent with the progressive loss of translational quasi–long-range order. Simultaneously, defect clusters emerge and rapidly grow, becoming the dominant source of structural disorder (see SM). Around $\varphi_2$, isolated disclinations are also detected, although their number remains small compared to that of defect clusters.

The two-step melting scenario observed here, consisting of a first-order liquid--hexatic transition followed by a continuous hexatic--solid transition, closely resembles the melting behavior of 2D thermal hard-sphere systems\cite{PhysRevE.87.042134,PhysRevLett.114.035702,PhysRevLett.118.158001,PhysRevLett.107.155704}. However, the present system is an out-of-equilibrium active matter system under homogeneous constant driving and thus lies beyond the scope of the KTHNY theory\cite{2dmelt-RevModPhys.60.161, JM-Kosterlitz_1973, Nelson-PhysRevB.19.2457, Young-PhysRevB.19.1855, nelson2002defects}. Recent numerical studies of 2D active Brownian disks have reported a similar melting scenario for particle activity characterized by a P\'{e}clet number $Pe<3$ \cite{2d_actdisk_PhysRevLett.121.098003, 2d_actdisk_digregorio2022unified}. The P\'{e}clet number $Pe$ measures the ratio of the persistence length to the particle size; in our system, $Pe<1$ \cite{chen2024anomalous}. Our results are therefore consistent with these studies. The microscopic origin of the first-order hexatic--fluid transition remains an open question, possibly related to the high defect density and extensive cluster nucleation observed in both systems \cite{2d_actdisk_digregorio2022unified}.

\subsection*{Outlook}
We report the experimental observation of a continuous spacetime crystalline state in a classical system of active granular disks in $2+1$ dimensions. The spontaneous breaking of spacetime translational symmetry is directly visible to the naked eye and manifests as the formation of a triangular periodic lattice, accompanied by the simultaneous onset of synchronized periodic motion that persists for up to one day and is remarkably robust to external noise.

Importantly, our experimental results reveal an intricate three-stage melting scenario in which spatial and temporal crystalline order melt at distinct critical values and through different physical mechanisms, supporting the independence of these spacetime symmetries (see visual summary in Fig.~\ref{fig5}). The delayed melting of the spatial crystalline component is consistent with the emergence of an intermediate hexatic phase, as expected in 2D systems.

Our findings establish the existence of exotic out-of-equilibrium crystalline phases of matter in spacetime and open new avenues for realizing macroscopic classical systems that exhibit complex spatiotemporal symmetry-breaking patterns.

\section*{Acknowledgments}
We thank Zi Cai and Yang Xu for collaboration at an early stage of this work.
We thank Maxim Chernodub, Julien Garaud, Antti Niemi, Frank Wilczek for useful discussions.

\section*{Methods}\label{app1methods}
\subsection*{Experimental Methods}
A schematic of the experimental setup is shown in Fig.~\ref{fig1}(A). The basic
microscopic units of the system are active granular particles of disk-shaped caps,
fabricated by 3D printing using a resin material. Each particle consists of a
disk-shaped ratchet cap with six staggered legs attached to its underside, as
illustrated in Fig.~\ref{fig1}(A). All particles are monodisperse, with tip diameter $9.7$ mm and pitch diameter
$D=8.8$~mm. The pitch diameter $D$ will be used as units on the length-scale. To improve stability under vertical vibration, a cylindrical post of
length $2$~mm and diameter $4$~mm is mounted beneath the ratchet cap, lowering
the particle’s center of mass. Compared to the active Brownian vibrators used in previous
studies~\cite{chen2024anomalous}, the present particles are smaller and feature fewer legs
(six instead of twelve). To compensate for the reduced leg number, the leg
thickness is increased to prevent mechanical failure during prolonged vibration.

A single layer of particles is placed on an aluminum plate and confined within a
circular boundary, which is mounted on an electromagnetic shaker
(Fig.~\ref{fig1}(A)). The shaker generates a vertical sinusoidal vibration at
frequency $100$~Hz along the $z$ direction, with a maximum acceleration of $3g$,
where $g\approx 9.81$ m/s$^2$ denotes the gravitational acceleration. A Basler CCD camera is positioned above the setup and continuously records particle configurations
for $20$~hours at a frame rate of $40$~frames per second.

During an experimental run, each particle independently absorbs energy from the
vertical vibration and dissipates it through inelastic collisions with
neighboring particles and with the vibrating plate. When energy injection and
dissipation balance on average, the system reaches a nonequilibrium steady state.
The observed collective dynamics result from the interplay between individual
particle activity and interparticle interactions.

\subsection*{Non-affine Motion}
For each particle $i$, a neighbor set $N_i$ is defined at time $t$. Local structural
rearrangements between configurations at times $t$ and $t+\Delta t$ are quantified
using the Falk--Langer non-affinity measure $D_{\min}^2$ \cite{PhysRevE.57.7192}. We introduce the
neighbor-relative vectors
\begin{equation}\label{eq:non-affine_motion_1}
\mathbf{R}_{ij}(t)=\mathbf{r}_{j}(t)-\mathbf{r}_{i}(t),\qquad
\mathbf{R}_{ij}(t+\Delta t)=\mathbf{r}_{j}(t+\Delta t)-\mathbf{r}_{i}(t+\Delta t)
\end{equation}

where $j\in N_i$ labels neighbors of particle $i$. The best-fit local affine
deformation gradient $\mathbf{J}_i$, which can be expressed as a linear transformation matrix, is obtained by minimizing the mean-square
residual
\begin{equation}\label{eq:non-affine_motion_2}
\mathbf{J}_i=\arg\min_{\mathbf{J}}\sum_{j\in N_i}
\left\|\mathbf{R}_{ij}(t+\Delta t)-\mathbf{J}\,\mathbf{R}_{ij}(t)\right\|^{2}
\end{equation}

The mean-square non-affine displacement is then defined as
\begin{equation}\label{eq:non-affine_motion_3}
 D_{\min,i}^{2}(t,\Delta t)=
\sum_{j\in N_i}
\left\|\mathbf{R}_{ij}(t+\Delta t)-\mathbf{J}_i\,\mathbf{R}_{ij}(t)\right\|^{2}
\end{equation}

Here, $\|\cdot\|$ denotes the Euclidean norm. We normalize the mean-square non-affine
displacement by the square of the particle diameter $D$ and use its logarithm, $\log_{10}\!\left( D_{\min}^2/D^2\right)$, as a kinetic descriptor.
\subsection*{Static and dynamic structure factors}
\subsubsection*{Static Structure Factor}
The 2D static structure factor is evaluated on a discrete polar grid in reciprocal space. The radial wavevectors $q$ are uniformly spaced with grid spacing $2\pi/L_{sys}$, where $L_{sys}\approx50~cm$ denotes the diameter of the system boundary. The azimuthal angle $\theta$ is uniformly sampled over the interval $[0,2\pi)$. These polar coordinates are mapped to Cartesian wavevectors as
\begin{equation}\label{static_structure_factor_1}
\mathbf{q} = (q \cos\theta, q \sin\theta),
\end{equation}
where $q=|\mathbf{q}|$.
For each time frame of the trajectory, only the $N$ particles located within the circular boundary are retained. The discrete spatial Fourier transform of the particle positions $\mathbf{r}_j$ is computed as
\begin{equation}\label{static_structure_factor_1}
    f(\mathbf{q}) = \sum_{j=1}^{N} \exp\!\left(-i \mathbf{q} \cdot \mathbf{r}_j\right)
\end{equation}
The instantaneous static structure factor for that frame is obtained from the squared modulus of this complex amplitude, normalized by the number of included particles,
\begin{equation}\label{static structure factor 2}
    S(\mathbf{q}) = \frac{1}{N} \left| f(\mathbf{q}) \right|^2
\end{equation}

\subsubsection*{Dynamic Structure Factor}

The longitudinal and transverse dynamical structure factors, \( S_L(q,\omega) \) and \( S_T(q,\omega) \), are computed via normal mode analysis. We consider a 2D system of \( N \) particles. The displacement of particle \( i \) from its time-averaged equilibrium position is defined as
\begin{equation}\label{dynamic_structure_factor_1}
\mathbf{u}_i(t) = \mathbf{r}_i(t) - \bar{\mathbf{r}}_i 
\end{equation}
Collect all displacements into a \(2N\)-dimensional vector \( \mathbf{U}(t) \), \\where $\mathbf{U}(t) =Col[\mathbf{u}_{1,x}(t), \mathbf{u}_{1,y}(t), \mathbf{u}_{2,x}(t), \mathbf{u}_{2,y}(t), …, \mathbf{u}_{N-1,x}(t), \mathbf{u}_{N-1,y}(t), \mathbf{u}_{N,x}(t), \mathbf{u}_{N,y}(t),]$, with `$Col[]$’ representing the column vector and $N$ being the total number of particles. The covariance matrix is then constructed by time averaging,
\begin{equation}\label{dynamic_structure_factor_2}
C_{\alpha\beta} = \langle U_\alpha(t)\, U_\beta(t) \rangle_t 
\end{equation}
The dynamical matrix is defined as the inverse of the covariance matrix,
\begin{equation}\label{dynamic_structure_factor_3}
D_M = C^{-1}.
\end{equation}
Diagonalization of \( D_M \) yields the normal modes,
\begin{equation}\label{dynamic_structure_factor_4}
D_M \mathbf{e}^m = \lambda_m \mathbf{e}^m 
\end{equation}
with eigenvalues \( \lambda_m \) and corresponding \(2N\)-dimensional eigenvectors \( \mathbf{e}^m \). The mode frequencies are given by
\begin{equation}\label{dynamic_structure_factor_5}
\omega_m = \sqrt{\lambda_m}
\end{equation}
Each eigenvector is projected into Fourier space as
\begin{equation}\label{dynamic_structure_factor_6}
\mathbf{\tilde U}(\mathbf{q},\omega_m) =
\sum_{i=1}^{N}
\mathbf{e}_i^m \,
e^{-i \mathbf{q} \cdot \bar{\mathbf{r}}_i}
\end{equation}
The resulting field is decomposed into longitudinal and transverse components,

\begin{equation}\label{dynamic_structure_factor_7}
\begin{aligned}
F_L(\mathbf{q},\omega_m) &=
\mathbf{q}\cdot \tilde{\mathbf{U}}(\mathbf{q},\omega_m), \\
F_T(\mathbf{q},\omega_m) &=
\left(\mathbf{q}\times \tilde{\mathbf{U}}(\mathbf{q},\omega_m)\right)\cdot \hat{\mathbf{z}} .
\end{aligned}
\end{equation}

The modes are grouped into frequency bins centered at \( \omega \). For each bin, the corresponding components are summed, normalized by \( q^2 \), and averaged over the azimuthal angle \( \theta_{\mathbf{q}} \),
\begin{equation}\label{dynamic_structure_factor_8}
S_{L,T}(q,\omega) =
\left\langle
\frac{1}{q^2}
\left|
\sum_{\omega_m \in \omega}
F_{L,T}(\mathbf{q},\omega_m)
\right|^2
\right\rangle_{\theta_{\mathbf{q}}}.
\end{equation}

The mode frequencies obtained from the covariance matrix method
have units of inverse length ($\text{length}^{-1}$).
To convert them into physical frequencies (inverse time),
they must be multiplied by a characteristic velocity scale
of the system.

In order to do so, the instantaneous velocity of each particle is computed
from the discrete trajectories as
\begin{equation}
\label{dynamic_structure_factor_9}
\mathbf{v}_i(t)
=
\frac{\mathbf{r}_i(t+\Delta t) - \mathbf{r}_i(t)}{\Delta t},
\end{equation}
where $\Delta t$ is the time interval between consecutive frames.

We define a characteristic kinetic speed of the active system as
the ensemble and time average of the absolute particle velocities:
\begin{equation}
\label{dynamic_structure_factor_10}
V_0
=
\left\langle
|\mathbf{v}_i(t)|
\right\rangle_{i,t}.
\end{equation}

The dimensionless mode frequencies $\omega_m$ are then converted
into physical angular frequencies (in units of $\text{s}^{-1}$)
via
\begin{equation}
\label{dynamic_structure_factor_11}
\omega_m^{\text{phys}}
=
V_0 \, \omega_m.
\end{equation}

This rescaling ensures that the vibrational spectrum
is expressed in physical time units.

\subsection*{Space Crystalline Fraction}
The static structure factor \( S(\mathbf{q}) \) is computed directly from the particle positions. 
To extract its angular dependence, we perform an angular averaging over the relevant 
wavevector shell and obtain \( S(q_\theta) \).

For a 2D triangular crystal, \( S(\mathbf{q}) \) displays six Bragg peaks 
arranged in a hexagonal pattern. The angular domain \( [0,2\pi) \) is therefore divided 
into six equal sectors:
\begin{equation}\label{Space_Crystalline_Fraction_1}
    [0,\pi/3),\; [\pi/3,2\pi/3),\; [2\pi/3,\pi),\; 
[\pi,4\pi/3),\; [4\pi/3,5\pi/3),\; [5\pi/3,2\pi)
\end{equation}

Within each sector, the peak position \( \theta_{p_i} \) of \( S(q_\theta) \) is identified. 
The spectral weight contained in an angular window of width \( l_W \) centered at each peak is then computed as
\begin{equation}\label{Space_Crystalline_Fraction_2}
    S_{p_i} = 
\int_{\theta_{p_i}-l_W}^{\theta_{p_i}+l_W}
S(q_\theta)\, d q_\theta
\qquad i=1,\ldots,6
\end{equation}

The angular grid consists of 360 uniformly spaced bins in the interval \( [0,2\pi] \). 
We choose \( l_W = 10^\circ \), which fully captures the width of the Bragg peaks.

The spatial crystalline fraction is defined as

\begin{equation}\label{Space_Crystalline_Fraction_3}
    f_{\text{space}} =
\frac{\sum_{i=1}^{6} S_{p_i}}{S_T}
\end{equation}

where
\begin{equation}\label{Space_Crystalline_Fraction_4}
   S_T =
\int_{0}^{2\pi} S(q_\theta)\, d q_\theta 
\end{equation}

is the total angularly integrated structure factor.

\subsection*{Time Crystalline Fraction}
The temporal Fourier transform of particle trajectories projected onto a chosen
axis (e.g., the $y$-axis) is computed. For a given particle trajectory within a time window $T_W$, the power spectrum $S(\omega)$ is defined as the squared modulus of its discrete Fourier transform normalized by $T_W$: 
\begin{equation}\label{Time_Crystalline_Fraction_1}
    S(\omega) = \frac{1}{T_W} \left| \sum_{t} y(t) e^{-i\omega t} \right|^2
\end{equation}

The dominant oscillation frequency $\omega_p$ is identified, and the spectral power within a frequency window $\Delta_W$ around this peak is summed:
\begin{equation}\label{Time_Crystalline_Fraction_2}
  S_{p,W} = \sideset{}{'}\sum_{\omega \in W_p} S(\omega)  
\end{equation}

where \( W_p := \{\omega \;|\; |\omega-\omega_p| \le \Delta_W \} \). In the discrete frequency domain, this window spans $n$ adjacent frequency bins on either side of the peak, such that $\Delta_W = n \delta \omega$, where $\delta \omega$ is the frequency resolution of the Fourier transform. The prime on the sum indicates that a low-frequency cutoff is applied (strictly excluding the zero-frequency DC component, \( \omega_0 = 0 \)), as it solely represents the mean position of the particle.

The time crystalline fraction is defined as:
\begin{equation}\label{Time_Crystalline_Fraction_3}
    f_{\text{time}} = \frac{S_{p,W}}{S_T}
\end{equation}

with $S_T = \sideset{}{'}\sum_{\omega} S(\omega)$ representing the total spectral power above the low-frequency cutoff. To ensure statistical robustness, the mean value and standard deviation of $f_{\text{time}}$ are obtained by computing and averaging this fraction over a random selection of independent particle trajectories.

\subsection*{Defect Identification}
Topological defects are identified via a Voronoi construction of the instantaneous 2D particle positions. 
In a perfect triangular lattice, each particle has coordination number $Z_i = 6$. 
Particles with $Z_i \neq 6$ are therefore classified as defects. Defects are characterized by two fundamental topological invariants: the Frank vector and the Burgers vector.

\medskip

\textbf{Frank Vector ($\Omega$).}
The Frank vector (or simply the Frank angle in 2D since its associated direction is fixed along $\mathbf{\hat{z}}$) quantifies rotational symmetry breaking. 
It is defined as the circulation of the local bond-orientational angle $\theta(\mathbf{r})$ around a closed contour $\Gamma$ enclosing the defect:
\begin{equation}\label{Defect_Identification_1}
    \Omega = \oint_{\Gamma} d\theta.
\end{equation}
In a triangular lattice with sixfold symmetry, the allowed disclination charges are quantized in units of $\pi/3$:
\begin{equation}\label{Defect_Identification_2}
    \Omega = \pm \frac{\pi}{3}, \pm \frac{2\pi}{3}, \dots
\end{equation}
The elementary defects correspond to 5-fold sites ($\Omega = +\pi/3$) and 7-fold sites ($\Omega = -\pi/3$), which represent isolated disclinations [Fig.~\ref{fig:defect}(a)]. 
These defects act as sources of rotational strain and generate long-range elastic distortions.

\medskip

\textbf{Burgers Vector ($\mathbf{b}$).}
The Burgers vector quantifies translational symmetry breaking. 
It is defined through the closure failure of a lattice circuit $\Gamma$ encircling the defect:
\begin{equation}\label{Defect_Identification_3}
    \mathbf{b} = \oint_{\Gamma} d\mathbf{u}
\end{equation}
where $\mathbf{u}(\mathbf{r})$ is the displacement field relative to a perfect reference lattice. 
A non-zero $\mathbf{b}$ corresponds to the termination of an extra half-plane of particles at the defect core.

In a triangular lattice, an elementary dislocation [Fig.~\ref{fig:defect}(b)] consists of a tightly bound 5--7 disclination pair. 
The opposite Frank charges cancel ($\sum \Omega = 0$), eliminating long-range rotational strain. 
However, the pair possesses a finite Burgers vector perpendicular to the separation vector $\mathbf{d}$ between the 5- and 7-fold sites:
\begin{equation}\label{Defect_Identification_4}
    \mathbf{b} = \Omega\, \hat{\mathbf{z}} \times \mathbf{d}
\end{equation}
where $\Omega = \pi/3$ is the elementary Frank angle. 
For the minimal dislocation, $|\mathbf{b}|$ equals one lattice spacing.

\medskip

\textbf{Dislocation Pairs (Quadrupoles).}
Two adjacent dislocations with opposite Burgers vectors form a quadrupolar defect [Fig.~\ref{fig:defect}(c)]. 
In this configuration both the net Frank charge and the net Burgers vector vanish:
\begin{equation}\label{Defect_Identification_5}
    \sum \Omega = 0, 
\qquad 
\sum \mathbf{b} = 0
\end{equation}

Such quadrupoles are not topological defects in the strict sense.

\medskip

Based on these invariants, defects are classified as follows:

\begin{itemize}
\item \textbf{Disclinations (isolated defects):} Single 5- or 7-fold particles carrying non-zero Frank charge ($\Omega \neq 0$).
\item \textbf{Dislocations (dipoles):} Bound 5--7 pairs with $\sum \Omega = 0$ but $\mathbf{b} \neq 0$.
\item \textbf{Quadrupoles:} Bound pairs of dislocations with $\sum \Omega = 0$ and $\sum \mathbf{b} = 0$.
\end{itemize}

\subsection*{Dynamical and Structural Percolation}
Dynamical percolation is quantified on the Voronoi neighbor graph. At each time frame
$t$, particles are classified as dynamically ordered if
$\log_{10}\!\big(D_{\min,i}^2(t)/D^2\big)<0$. Ordered particles are connected using
Voronoi adjacency, yielding a set of clusters $\mathcal{C}$. The geometric extent
of a cluster is defined as
\begin{equation}\label{Dynamical_and Structural_Percolation_1}
    L_{\mathrm{dyn}}(t)=\max_{i,j\in\mathcal{C}}
\left\|\mathbf{r}_i(t)-\mathbf{r}_j(t)\right\|
\end{equation}

The system size $L_{\mathrm{sys}}$ is taken as the diameter of the circular
boundary. A frame is considered dynamically percolated if the largest ordered
cluster spans the system,
$\max_{\mathcal{C}} L_{\mathrm{dyn}}(t) =  L_{\mathrm{sys}}$. The dynamical
percolation probability is then
\begin{equation}\label{Dynamical_and Structural_Percolation_2}
    P_{\mathrm{dyn}}=\frac{1}{N_{\mathrm{f}}}\sum_{t=1}^{N_{\mathrm{f}}}
H\!\left(\max_{\mathcal{C}} L_{\mathrm{dyn}}(t)-L_{\mathrm{sys}}\right)
\end{equation}

where $N_{\mathrm{f}}$ is the number of time frames analyzed and $H(x)$ is the Heaviside
step function.

Structural percolation is defined analogously, with structurally ordered particles
identified by the hexatic order parameter (as will be defined in the later section), $|\psi_6(i,t)|>0.64$. Clusters are
constructed using the same Voronoi adjacency, and the structural percolation
probability reads
\begin{equation}\label{Dynamical_and Structural_Percolation_3}
    P_{\mathrm{str}}=\frac{1}{N_{\mathrm{f}}}\sum_{t=1}^{N_{\mathrm{f}}}
H\!\left(\max_{\mathcal{C}_6} L_{\mathrm{str}}(t)-L_{\mathrm{sys}}\right)
\end{equation}

\subsection*{Directional Persistence}
Long-time correlations of velocity directions in the co-rotating frame of the time
crystal are quantified using a directional persistence measure. For each particle
$i$, the instantaneous velocity direction
$\hat{\mathbf{v}}_i(t)=\mathbf{v}_i(t)/\|\mathbf{v}_i(t)\|$ is compared with the local
tangential direction $\hat{\mathbf{t}}_i(t)$, defining the misalignment angle

\begin{equation}\label{Directional_Persistence_1}
    \theta_i(t)=\arccos\!\bigl(\hat{\mathbf{v}}_i(t)\cdot \hat{\mathbf{t}}_i(t)\bigr)
\end{equation}

The single-particle persistence at lag time $\Delta t$ is
\begin{equation}\label{Directional_Persistence_2}
    P_i(\Delta t)=\left\langle
\cos\!\left[\theta_i(t_0+\Delta t)-\theta_i(t_0)\right]
\right\rangle_{t_0}
\end{equation}

where the average is taken over time origins $t_0$ in the steady state. The global
persistence is obtained by averaging over particles,
\begin{equation}\label{Directional_Persistence_3}
    P(\Delta t)=\frac{1}{N}\sum_{i=1}^N P_i(\Delta t)
\end{equation}

The lag time is chosen as a packing-fraction-dependent characteristic time
$t^\ast$, defined from the normalized mean-squared displacement
$\mathrm{MSD}(t)/D^2$ via $\mathrm{MSD}(t^\ast)/D^2=2.5$. Directional persistence is
evaluated at $\Delta t=t^\ast$.

\subsection*{Correlation functions}
\subsubsection*{Translational correlation function}
To calculate the translational correlation function $G_T(r)$, the primary reciprocal lattice vector $\mathbf{G}$ is first identified by locating the highest peak of the static structure factor $S(\mathbf{q})$ within the first Brillouin zone. Using $\mathbf{G}$, a local translational order parameter is defined for each particle $j$ at position $\mathbf{r}_j$:
\begin{equation}\label{Correlation_functions_1}
    \psi_T(\mathbf{r}_j) = \exp(i \mathbf{G} \cdot \mathbf{r}_j)
\end{equation}

To evaluate how well this positional order persists in space, the spatial autocorrelation of $\psi_T$ is computed. Random center particles, positioned sufficiently far from the system boundaries to avoid edge effects, are selected, and the complex product with their neighbors $k$ is evaluated:
\begin{equation}\label{Correlation_functions_2}
    \psi_T^*(\mathbf{r}_j)\psi_T(\mathbf{r}_k)
\end{equation}

where $r = |\mathbf{r}_k - \mathbf{r}_j|$ is the separation distance between the pair. Finally, these pairwise products are grouped into radial bins based on $r$. The translational correlation function $G_T(r)$ is defined explicitly as the ensemble average of these products:
\begin{equation}\label{Correlation_functions_3}
    G_T(r) = \langle \psi_T^*(\mathbf{r}_j) \psi_T(\mathbf{r}_k) \rangle_{|\mathbf{r}_k - \mathbf{r}_j| = r}
\end{equation}

where the angle brackets indicate an average taken across all sampled pairs at distance $r$ and across multiple time frames.
\subsubsection*{Sixfold bond-orientational correlation function}
Sixfold bond-orientational order is characterized through the hexatic order parameter defined over Voronoi nearest neighbors. The local complex order parameter for particle $i$ is given by
\begin{equation}\label{Correlation_functions_4}
    \psi_6(i)=\frac{1}{N_i}\sum_{j\in N_i} 
e^{\,\mathrm{i}\,6\,\theta_{ij}}
\end{equation}

where $N_i$ denotes the set of nearest neighbors of particle $i$. 
For each $j \in N_i$, $\theta_{ij}$ is the angle between the bond vector 
$\mathbf{r}_{ij} = \mathbf{r}_j - \mathbf{r}_i$ and a fixed reference axis (e.g., the $x$-axis).

The global hexatic order parameter is defined as
\begin{equation}\label{Correlation_functions_5}
    \Psi_6=\left|\frac{1}{N}\sum_{i=1}^{N}\psi_6(i)\right|
\end{equation}

where $N$ is the total number of particles.

\medskip
Temporal correlations of the bond-orientational order are quantified by the normalized autocorrelation function
\begin{equation}\label{Correlation_functions_6}
    G_6(t)=
\frac{
\left\langle 
\psi_6(i,t_0+t)\,\psi_6^{\ast}(i,t_0)
\right\rangle_{i,t_0}
}{
\left\langle 
\left|\psi_6(i,t_0)\right|^2
\right\rangle_{i,t_0}
}
\end{equation}

Here, $i$ labels the particle index, $t$ is the time lag, and $t_0$ is the time origin. 
The numerator correlates the complex hexatic order parameter of particle $i$ at time $t_0$ with its value at the later time $t_0+t$. 
The superscript ${}^{\ast}$ denotes complex conjugation. 
The average $\langle \cdots \rangle_{i,t_0}$ is taken over all particles $i$ and over all admissible time origins $t_0$ such that both $t_0$ and $t_0+t$ lie within the recorded time window. 
The denominator provides normalization, ensuring $G_6(0)=1$ and removing dependence on the overall magnitude of $\psi_6$.

\medskip
Spatial correlations are characterized by

\begin{equation}\label{Correlation_functions_7}
    G_6(r)=
\frac{
\left\langle 
\psi_6(i,t)\,\psi_6^{\ast}(j,t)
\right\rangle_{
|{\bf r}_i-{\bf r}_j|\in[r,r+\Delta r],\,t
}
}{
\left\langle 
|\psi_6(i,t)|^2
\right\rangle_{i,t}
}
\end{equation}

Here, $r$ denotes the spatial separation and $\Delta r$ is the radial bin width. 
The vectors ${\bf r}_i(t)$ and ${\bf r}_j(t)$ represent the particle positions at time $t$, and $|{\bf r}_i-{\bf r}_j|$ is their center-to-center distance. 
The numerator averages over all particle pairs $(i,j)$ whose separation lies within the interval $[r,r+\Delta r]$, and additionally over all recorded time frames. 
The denominator averages $|\psi_6(i,t)|^2$ over all particles and times, ensuring that $G_6(r)$ is dimensionless and independent of the overall magnitude of $\psi_6$.

\subsubsection*{Projected temporal autocorrelation function}
The time auto-correlation function $G(t)$ of the particles' periodic motion is calculated from their positional projections along a specified axis (e.g., the $y$-axis). The raw coordinate projection $y_i(t)$ of particle $i$ is first transformed into a normalized signal $\tilde{y}_i(t)$ strictly bounded within $[-1, 1]$ using min-max scaling:
\begin{equation}\label{Correlation_functions_8}
    \tilde{y}_i(t) = 2 \left( \frac{y_i(t) - \min_t(y_i(t))}{\max_t(y_i(t)) - \min_t(y_i(t))} \right) - 1
\end{equation}
This bounds the projection range to a normalized $[-1, 1]$ interval. The time auto-correlation of these normalized projections is then computed as:
\begin{equation}\label{Correlation_functions_9}
    G(t) = \langle \tilde{y}_i(\tau) \tilde{y}_i(t+\tau) \rangle_{\tau, i}
\end{equation}
Here, $t$ is the lag time representing the delay between two observations, $\tau$ denotes the time origins, and $i$ specifies the particle labels. The angle brackets $\langle \dots \rangle_{\tau, i}$ indicate a double averaging process: an inner time-average taken over the various time origins $\tau$, followed by an outer ensemble average taken over the valid particles $i$ in the system. 

In the computational implementation, particles initially lying inside an annulus of radius $8.8\text{ cm} < R < 20.6\text{ cm}$ are selected for the ensemble average. For a given time lag $t$, $20$ time origins $\tau$ are randomly sampled from a uniform distribution over the valid interval $[0, T-t]$ independently for each particle, where $T$ is the total time recorded in the experiment.

Moreover, $t_{\max}$ is truncated to guarantee that an adequate number of independent time origins remains available for sampling at large time lags. The power-law decay exponent is extracted by identifying the local maxima (the true envelope) of $|G(t)|$ and performing a linear regression in log-log space. The random origin sampling to the envelope peak-fitting is independently repeated 40 times. The uncertainty is the standard deviation of the slopes.

\section*{Data availability}
The datasets generated and analyzed during the current study are available upon reasonable request by contacting the corresponding authors. 

\section*{Funding}
JB and MB acknowledge the support of the Shanghai Municipal Science and Technology Major Project (Grant No.2019SHZDZX01).
GL and JZ acknowledge the support of the NSFC (Nos. 12534008, 11974238, and 12274291) and also the support of the Innovation Program of Shanghai Municipal Education Commission under No. 2021-01-07-00-02-E00138.

\section*{Author contributions statement}
JZ and MB conceived the idea of this work and supervised it. GL performed the experiments and conducted the data analysis together with JB. All the authors contributed to the preparation of the manuscript and the discussion and interpretation of the results.

\clearpage
\newpage

\begin{figure}[H]
    \centering
    \includegraphics[width=\textwidth]{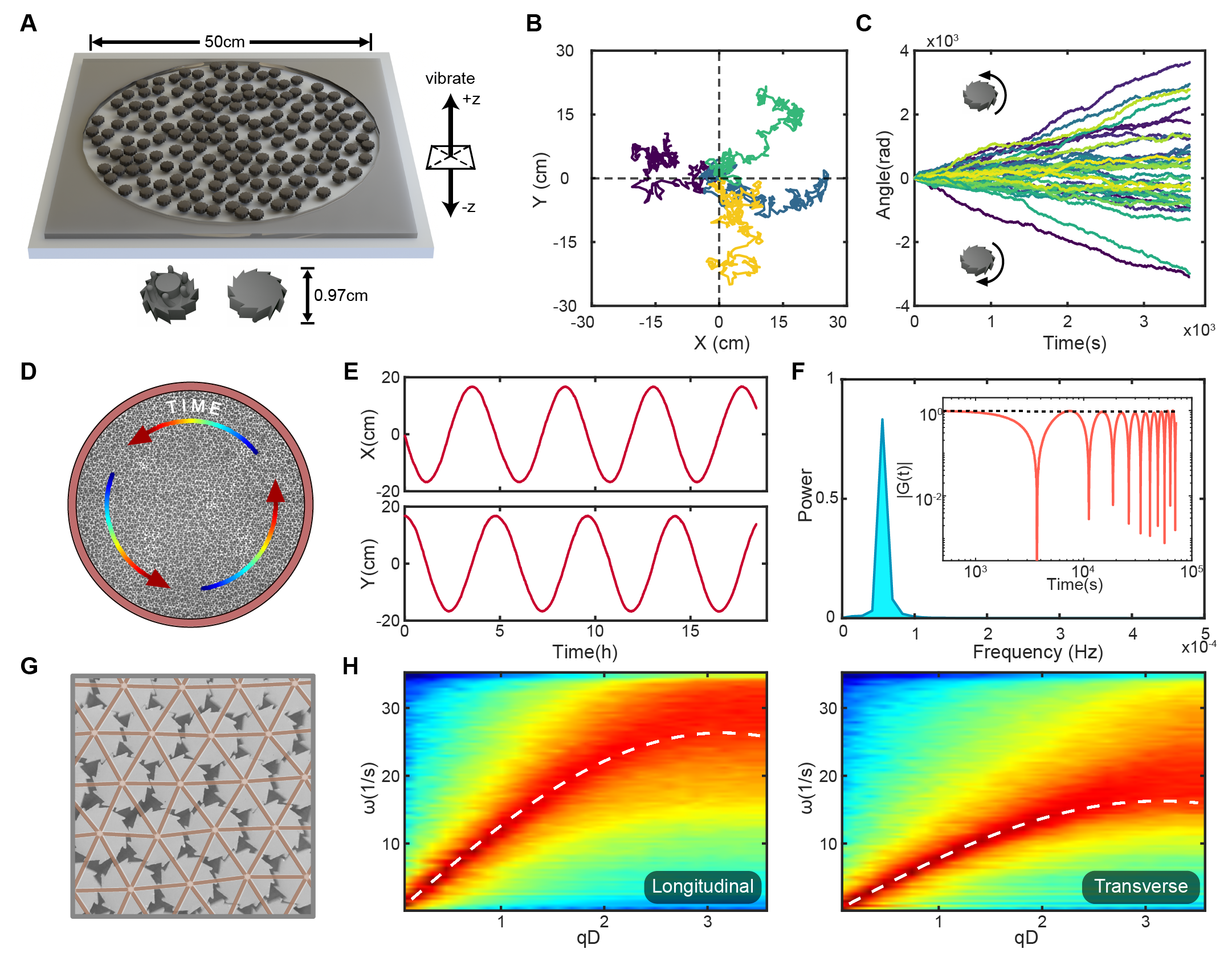} 
    \caption{\textbf{Macroscopic continuous spacetime crystal}.
    \textbf{A.} Experimental setup: Particles are placed on an aluminum-alloy plate and confined within a circular boundary. Vertical vibrations are applied normal to the plate using an electromagnetic shaker. Inset: schematic of the ratcheted particles.
\textbf{B.} Single-particle translational motion: Particles initially placed near the center exhibit random, disordered trajectories, consistent with 2D diffusive Brownian motion.
\textbf{C.} Single-particle rotational motion: External vibrations apply torque, although most spin angular momenta remain near zero. Theoretically, a particle has zero mean spin due to the symmetric inclination of its six legs, whereas a disk acquires finite spin from manufacturing imperfections.
\textbf{D.} Photograph of the macroscopic time crystal. Trajectories of three representative particles are shown. The color gradient from blue to red indicates particle positions over a time interval from $0$ to $2400$~s. The particles move collectively, forming a global spacetime crystalline state, as shown in Supplementary Movie~\ref{vid:835}.
\textbf{E.} The $X$ and $Y$ position of a benchmark particle in panel D as a function of time at $\varphi = 0.835$.
\textbf{F.} Power spectrum of the phase signal. A pronounced peak appears at $f = 5.5 \times 10^{-5}\,\mathrm{Hz}$ corresponding to a rotation period of $5.05$~h. Inset: time-correlation function of the normalized projection of particles' motion $\tilde{y}$ in log--log scale, defined as $G(t)=\langle \tilde{y}(t)\tilde{y}(0)\rangle - \langle\tilde{y}(t)\rangle\langle\tilde{y}(0)\rangle$, where the ensemble average is taken over particles of radius $8.8\text{cm}<R<20.6\text{cm}$ to avoid center and boundary region. The black dashed line shows a fit to the envelope of |$G(t)$|, yielding a slope of $-0.09\pm0.05$. \textbf{G.} Magnified photograph of a portion of the spacetime crystal. Particles arrange into a triangular lattice.
\textbf{H.} Longitudinal and transverse dynamical structure factors $S(q,\omega)$ after subtraction of the global rotation. Here $D$ denotes the pitch diameter ($D=8.8$~mm). The white dashed lines show fits to the maxima of $S(q,\omega)$, given by $\omega=26.34\sin(qD/2)$ for the longitudinal mode and $\omega=16.27\sin(qD/2)$ for the transverse mode, where both frequencies are measured in Hz (1/s).}
    \label{fig1}
\end{figure}

\begin{figure}[H]
    \centering
    \includegraphics[width=\textwidth]{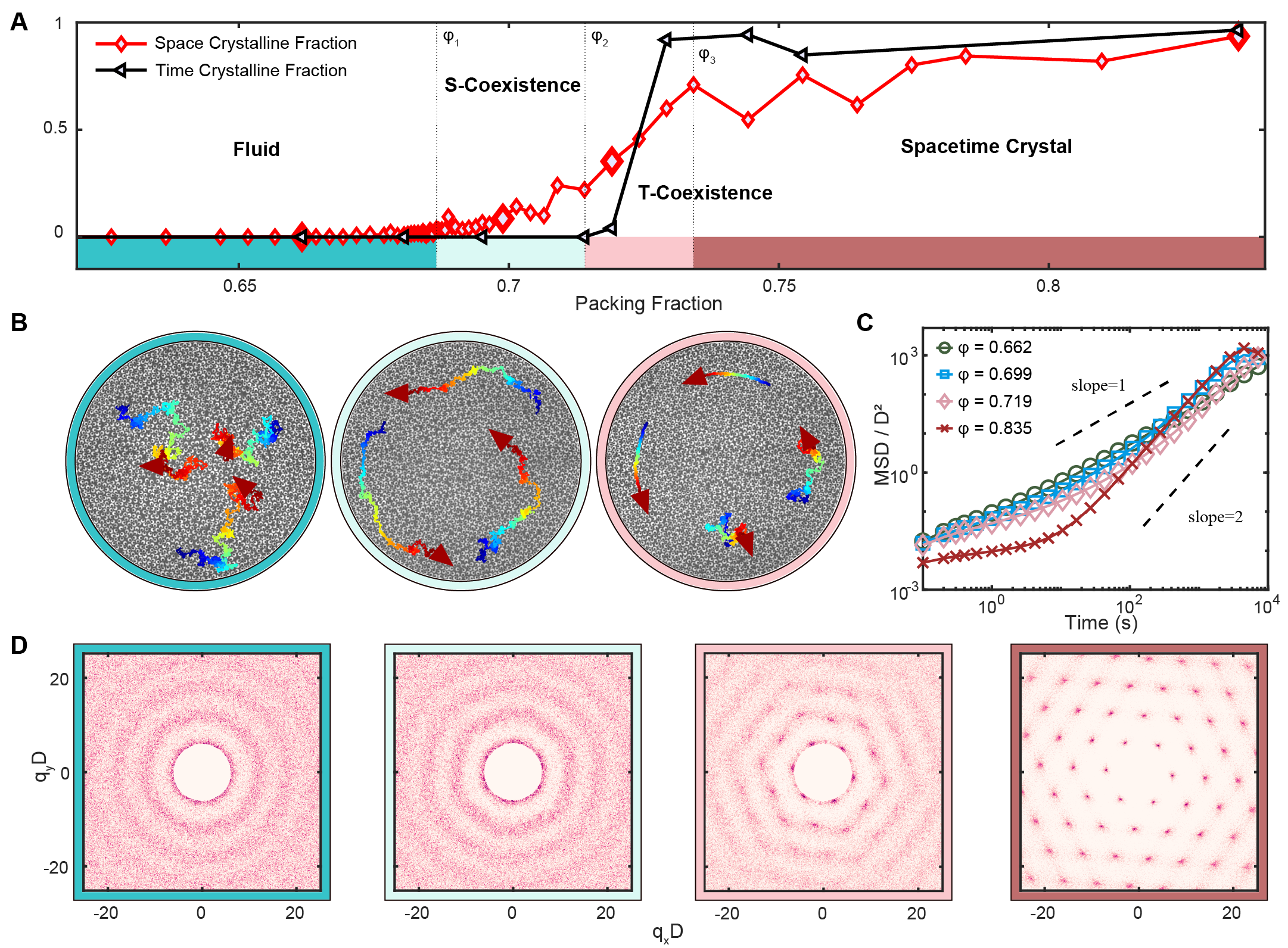} 
    \caption{\textbf{Three-stage melting of spacetime order}.
   \textbf{A.} Phase diagram showing, from right to left, the spacetime crystal, time-coexistence, space-coexistence, and fluid phases. The temporal and spatial crystalline fractions decrease independently across the coexistence regimes. The critical packing fractions separating the phases are $\varphi_1 = 0.687$, $\varphi_2 = 0.709$, and $\varphi_3 = 0.734$. One representative packing fraction is selected for each phase to illustrate its characteristic behavior; in ascending order, these are $\varphi = 0.662$, $0.699$, $0.719$, and $0.835$. The corresponding locations are indicated by enlarged red diamonds.
\textbf{B.} Particle trajectories at three selected packing fractions, as shown in Supplementary Movies~\ref{vid:662}, \ref{vid:699}, and~\ref{vid:719}. The color scale, ranging from blue to red, represents particle positions over the time interval \(0\) to \(2400\)\,s. Left: random trajectories. Middle: weak dynamical coherence. Right: coexistence of clustered and disordered particle motion.
\textbf{C.} Normalized mean-square displacements (MSD) at the four selected packing fractions. At $\varphi = 0.662$, the MSD displays diffusive scaling (slope $=1$), whereas at $\varphi = 0.835$ it exhibits ballistic scaling associated with rigid-body rotation (slope $=2$). For clarity, the same color code is used throughout the figure to denote these four packing fractions.
\textbf{D.} Spatial patterns of the static structure factor $S(q_xD,q_yD)$. As the packing fraction decreases from right to left, the system evolves from a crystalline state to a hexatic phase, then melts, and ultimately transitions into a fluid.}
    \label{fig2}
\end{figure}

\begin{figure}[H]
    \centering
    \includegraphics[width=\textwidth]{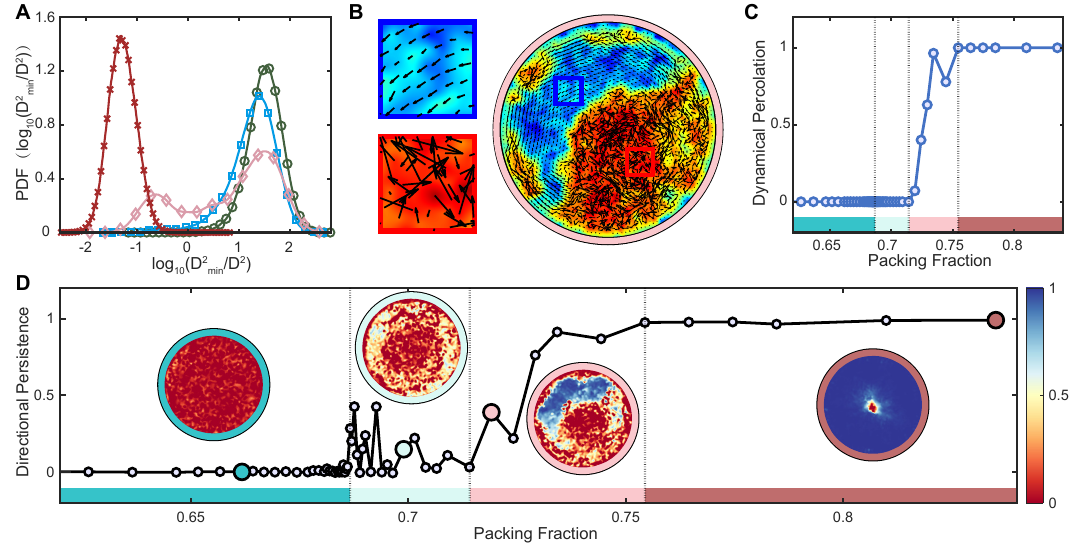} 
    \caption{\textbf{Dynamical melting of time-crystalline order}.
    \textbf{A}. Dynamical coexistence. The dynamical parameter $\log_{10}\!\left(D_{\min}^2/D^2\right)$ quantifies local non-affine displacements. In the spacetime crystalline phase, this quantity remains small, indicating highly ordered and coherent dynamics. In the fluid phase, no ordered domains are present. In the time-coexistence (T-coexistence) region, dynamically ordered and disordered domains coexist. \textbf{B}. Schematic of the coexistence between a time crystal and a fluid. Arrows indicate particle displacements over $200\,\mathrm{s}$ after rescaling. Color encodes the magnitude of the non-affine displacement (blue: small; red: large). \textbf{C}. Percolation of dynamical order. The spatial extent of the time-crystalline domain decreases rapidly across the T-coexistence regime. \textbf{D}. Directional persistence, which quantifies long-time correlations of velocity directions in the co-rotating frame of the time crystal. From right to left: \emph{Spacetime Crystal}, where each particle maintains alignment with its initial direction of motion; \emph{T-coexistence}, where the ability to sustain coherent tangential motion rapidly weakens as the packing fraction decreases; \emph{S-coexistence}, where particles largely lose directional persistence, with only sporadic tangential motion; \emph{Fluid}, where long-time velocity-direction correlations vanish completely. 
Inset: spatial map of the local directional persistence, where colors from blue to red denote long-time velocity-direction correlations from fully aligned to completely random. The red region near the center in the spacetime crystalline phase reflects deviations from perfect rigid-body rotation.}
    \label{fig3}
\end{figure}

\begin{figure}[H]
    \centering
    \includegraphics[width=\textwidth]{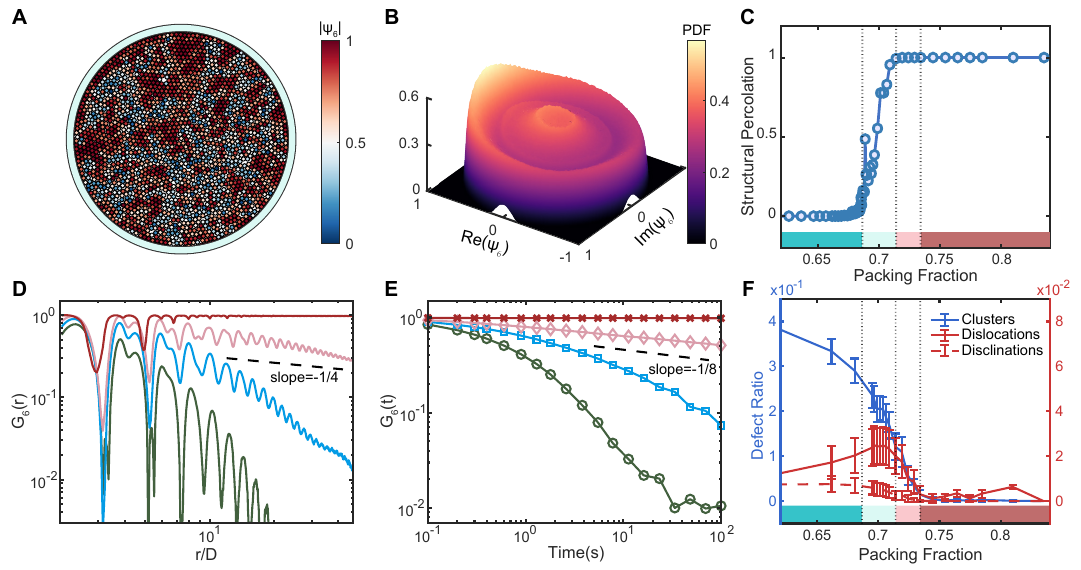} 
    \caption{\textbf{Structural correlations and topological defects}.
    \textbf{A}. Spatial distribution of the local hexatic order parameter $|\psi_6|$ in the S-coexistence phase at $\varphi=0.662$, where fragmented ordered domains coexist with a disordered fluid. Here, $|\psi_6|$ represents the modulus of $\psi_6$.
\textbf{B}. Distribution of $\psi_6$ in the complex plane for the configuration shown in panel A. The central peak corresponds to disordered fluid particles, while the peripheral peaks indicate locally ordered domains.
\textbf{C}. Percolation of structural order. The spatial extent of the space-crystalline domains decreases rapidly upon entering the S-coexistence regime.
\textbf{D}. Spatial hexatic correlation function. Three distinct decay regimes are observed: constant (Spacetime Crystal), power-law (T-coexistence and S-coexistence), and exponential (Fluid). The black dashed line with slope $-1/4$ marks the critical decay separating the hexatic-like and fluid phases.
\textbf{E}. Temporal hexatic correlation function. The black dashed line with slope $-1/8$ again indicates the boundary between the hexatic-like and fluid phases.
\textbf{F}. Defect ratio, defined as the ratio between the number of defects and the total number of particles in each frame, shown as a function of $\varphi$. Free dislocations proliferate at $\varphi=0.734$, signaling the loss of spatial long-range order and the transition from the crystalline to the hexatic-like phase. Free disclinations proliferate at $\varphi=0.714$, indicating the loss of angular order and the onset of the fluid phase. Defect clusters proliferate throughout the hexatic-like and coexistence phases.  }
    \label{fig4}
\end{figure}

\begin{figure}[H]
    \centering
    \includegraphics[width=\textwidth]{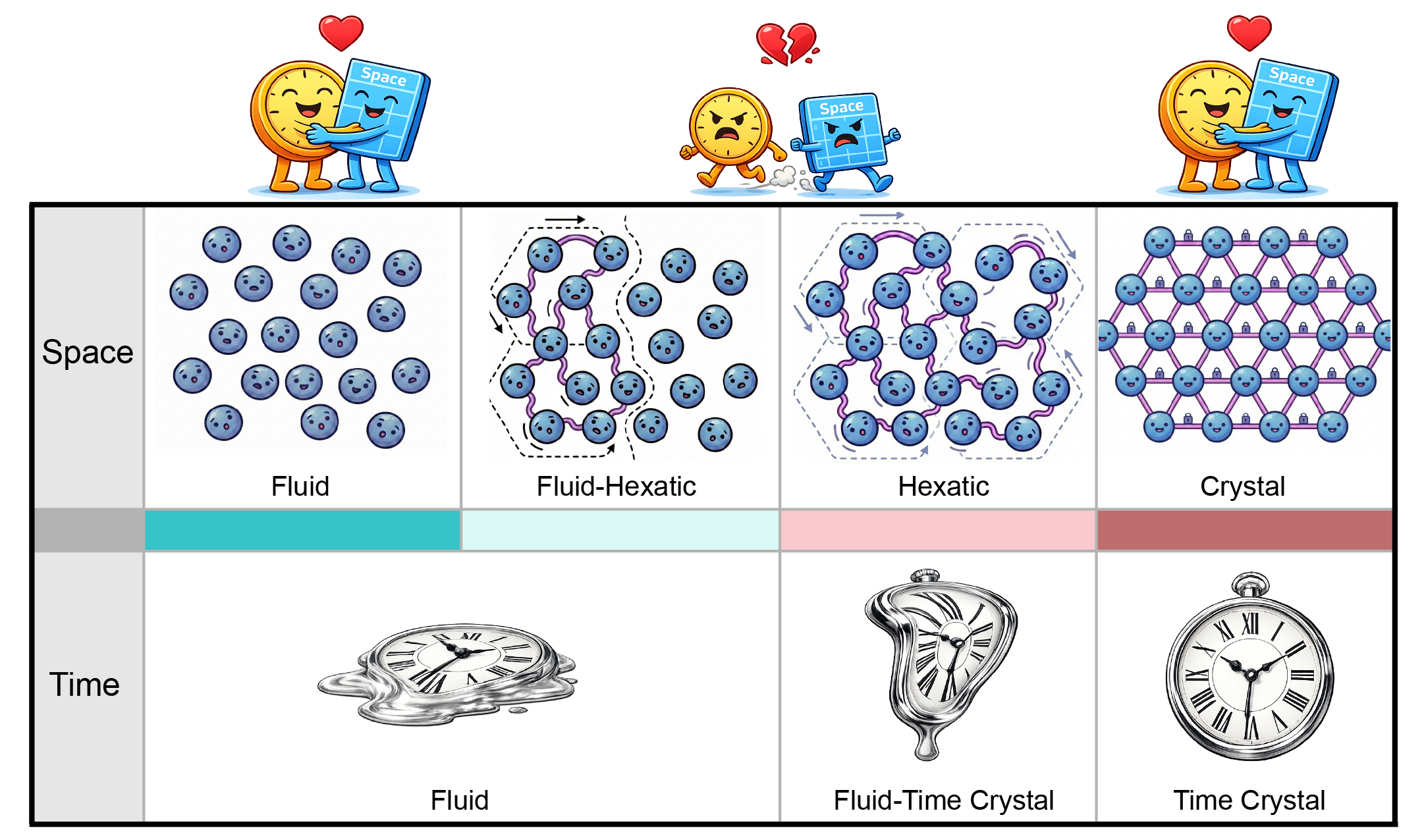} 
    \caption{\textbf{Spatiotemporal phase diagram of spatial and temporal order}.
A summary of the different phases in terms of their spatial and temporal crystalline order is shown. Remarkably, spatial and temporal symmetries are either both preserved or both broken in the low-packing-fraction fluid phase and in the high-packing-fraction spacetime crystalline state. In contrast, in the two intermediate coexistence regions, spatial and temporal order decouple: they melt at distinct critical points and through fundamentally different physical mechanisms, producing an intricate three-stage melting scenario including a hexatic state.}
    \label{fig5}
\end{figure}

\newpage
\clearpage

\appendix 
\renewcommand\thefigure{S\arabic{figure}}    
\setcounter{figure}{0} 
\renewcommand{\theequation}{S\arabic{equation}}
\setcounter{equation}{0}
\renewcommand{\thesubsection}{SM\arabic{subsection}.}
\clearpage
\newpage
\section*{\Large Supplementary Materials}

\setcounter{equation}{0}
\renewcommand{\theequation}{S\arabic{equation}} 

\subsection{Spontaneous breaking of time translations and random phase}
The results presented in the main text correspond to the primary experimental setup, consisting of a single large system with a diameter of approximately $50$ cm (see Fig.~\ref{fig1}A).

\begin{figure}
    \centering
    \includegraphics[height=6.25cm]{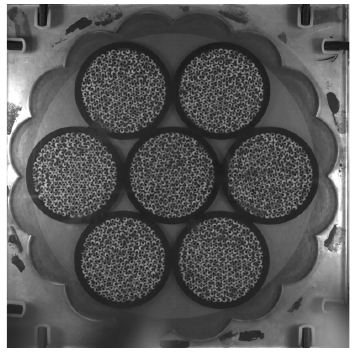}
    \includegraphics[height=6.25cm]{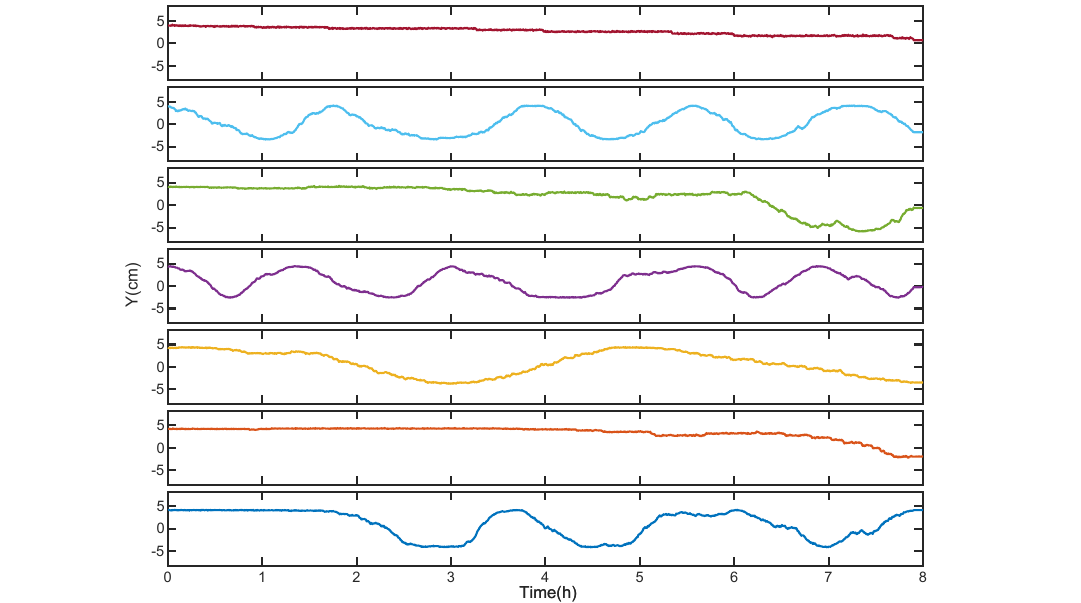}
    \caption{\textbf{A.} Snapshot of the experimental setup used to verify the random phase of the time-crystalline state. Seven subsystems, each $15$ cm in diameter, are mounted on an electromagnetic shaker and driven simultaneously. See Supplementary Movie~\ref{vid:random_phase} for a recording of the corresponding real-time dynamics. \textbf{B.} Time evolution of the $Y$-coordinate of a particle located at the same spatial position in seven independently prepared but otherwise identical systems, as shown in Fig.~\ref{SM1}. The resulting periodic motions are clearly out of phase, a feature that is confirmed systematically across multiple realizations of the time-crystalline state.}
    \label{SM1}
\end{figure}

To further substantiate the spontaneous breaking of time-translation symmetry, we must demonstrate that the phase of the emergent periodic motion is random, as expected from the canonical Mexican-hat scenario. To this end, we designed a second experimental configuration (see Fig.~\ref{SM1}A), comprising seven smaller systems, each with a diameter of $15$ cm, vibrated simultaneously. These systems are placed on top of the same shaker and can be regarded as seven identical yet independent replicas.

Following an initial quiescent interval, whose duration varies across replicas, each system begins to rotate. The onset times differ from one replica to another, with additional systems progressively entering the rotating state over time (see Supplementary Movie~\ref{vid:random_phase}). 

These rotating systems therefore constitute independent realizations of the time-crystalline phase, whose dynamics are shown in Fig.~\ref{SM1}B. Within the $15$ h experimental window, five out of the seven replicas begin rotating, exhibiting rotation periods of $2.06$ h, $4.95$ h, $1.68$ h, $1.85$ h, and $2.78$ h, respectively. The small variations in the measured periods are likely attributable to finite-size effects. In contrast, repeated realizations of the large ($50$ cm) system consistently yield the same period. Importantly, even when two replicas exhibit similar rotation periods, their phases remain uncorrelated. The remaining two replicas display onset times beyond the duration of the experimental observation window.

The stochastic onset of periodic motion and the randomness of the corresponding phases across independent realizations provide clear evidence of spontaneous time-translation symmetry breaking.

\subsection{Stability of the time crystalline phase upon external noise}

Rigidity and robustness against external perturbations are key criteria for identifying a bona fide time-crystalline state.

To test this requirement, we mounted a small system on a bone-conduction speaker that serves as a controlled noise source. The speaker generates non-sinusoidal vibrations, with the noise amplitude tuned via the volume control. It is placed directly on the electromagnetic shaker, allowing us to probe the system’s resilience under externally imposed noise. We refer to Supplementary Movie~\ref{vid:noise} for an example of the effects of the external noise on our experimental system.

We define the \emph{noise strength parameter}
\begin{equation}
\eta \equiv 
\frac{E_{\mathrm{speaker}}}{E_{\mathrm{vibrator}}}
\end{equation}
where $E_{\mathrm{speaker}}$ denotes the average particle translational kinetic energy in the presence of noise, and $E_{\mathrm{vibrator}}$ is the corresponding baseline value measured with the speaker turned off.

\begin{figure}
    \centering
    \includegraphics[width=0.6\linewidth]{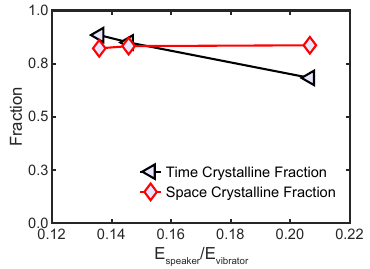}
    \caption{Space and time crystalline fractions in a small system at $\varphi=0.806$ as a function of the noise level.}
    \label{SM2}
\end{figure}

Under these conditions, we compute both the spatial and temporal crystalline fractions as functions of the noise level $\eta$. For the highest packing fraction $\varphi=0.835$, we do not notice any visible change in these quantities upon increasing the noise level, confirming a very high degree of stability of our spacetime crystal upon external perturbations. On the other hand, for a high but not maximal packing fraction $\varphi=0.806$, the results are reported in Fig.~\ref{SM2}. The spatial crystalline fraction remains entirely insensitive to external noise. In contrast, the time-crystalline fraction decreases only gradually as $\eta$ increases, remaining as high as $\sim 0.75$ even at the maximum achievable noise level.

In summary, the system preserves strong spatial and temporal order under externally applied noise, demonstrating that the time-crystalline state is robust and resilient to perturbations.

\subsection{Results of different types of particles}

By modifying the structural design of the particle cap and legs at a packing fraction of $\varphi = 0.835$, we investigate how particle geometry affects both the rotation direction and the crystalline properties.

\begin{figure}
    \centering
    \includegraphics{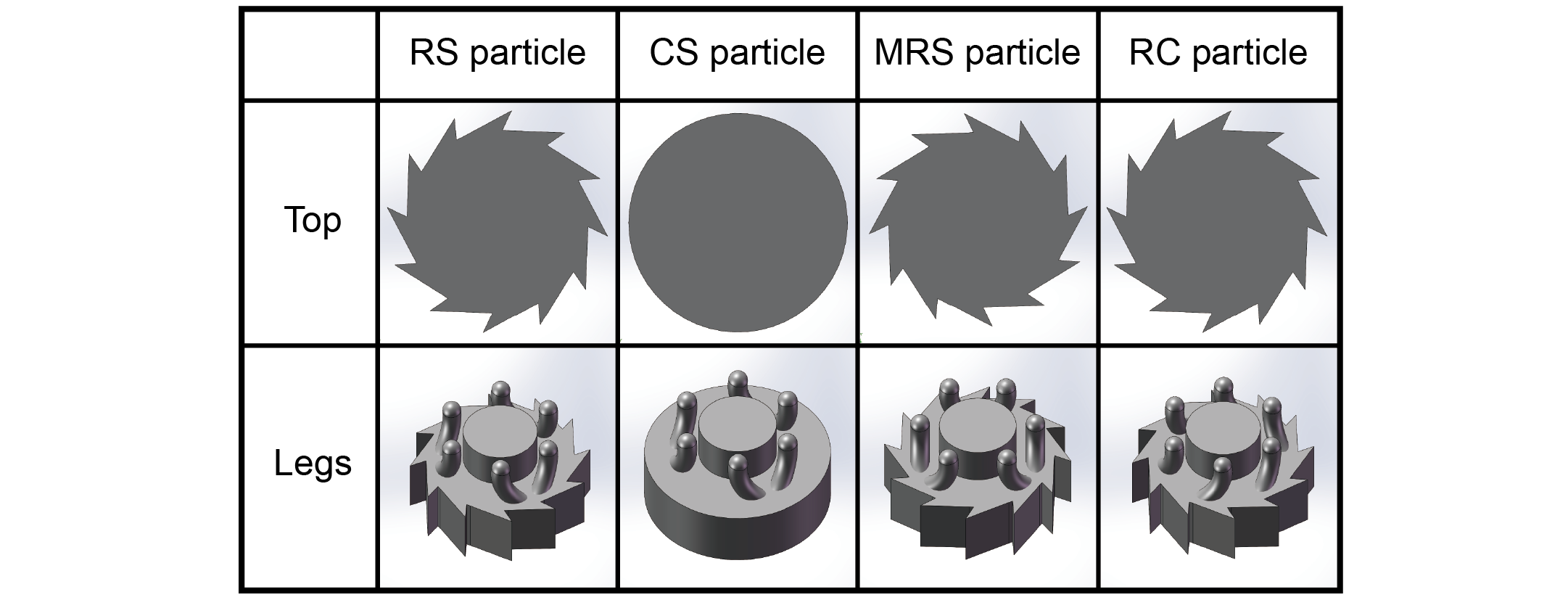} 
    \caption{
Images of the four particle types. The tip diameter of all types of ratchet particles (RS, MRS, and RC) is $9.7,\mathrm{mm}$ and the pitch diameter is $8.8,\mathrm{mm}$, while the diameter of the circular cap in the CS particle is $8.8,\mathrm{mm}$. The positions of the legs and the cylinder used to lower the center of mass are identical for all particle types.
}
    \label{SM_particle_types}
\end{figure}

We consider four particle types: RS, MRS, CS, and RC (see Fig.~\ref{SM_particle_types}). The RS particles consist of a ratchet cap and legs with alternating tilt, resulting in no net structural chirality. All the results in the main text refer to this type of particles. The MRS particles are identical to the RS particles except for a mirror-image ratchet cap. The CS particles are also identical to RS, but with a smooth disk cap instead of a ratchet. Finally, the RC particles differ from RS in that all legs tilt in the same direction, thereby introducing structural chirality.

At $\varphi = 0.835$, deep inside the spacetime-crystal phase, RS, MRS, and CS particles all rotate counterclockwise (see Supplementary Movies {~\ref{vid:835},~\ref{vid:MRS}~\ref{vid:CS}}). This common rotation direction indicates that the chirality of the time-crystalline motion is dictated by a small external bias present in the experimental setup, rather than by the specific particle geometry.

RC particles constitute an exception, as shown in Supplementary Movie~\ref{vid:RC}. For this geometry, the rotation direction is entirely determined by the intrinsic chirality of the legs and is therefore insensitive to residual experimental biases.

We stress that such experimental biases are negligible for packing fractions $\varphi < 0.729$, as discussed in the main text. As $\varphi$ decreases from its maximum value, the mean-square particle velocity increases monotonically and rapidly, as previously reported in~\cite{jiang2025experimental}. This increase effectively raises the kinetic temperature of the system, suppressing the influence of weak external biases.

Finally, although the chirality of the time-crystalline motion is largely set by external biases, the breaking of time-translation symmetry itself remains spontaneous.

\subsection{Extended data of particle trajectories}
\begin{figure}
    \centering
    \includegraphics[width=\textwidth]{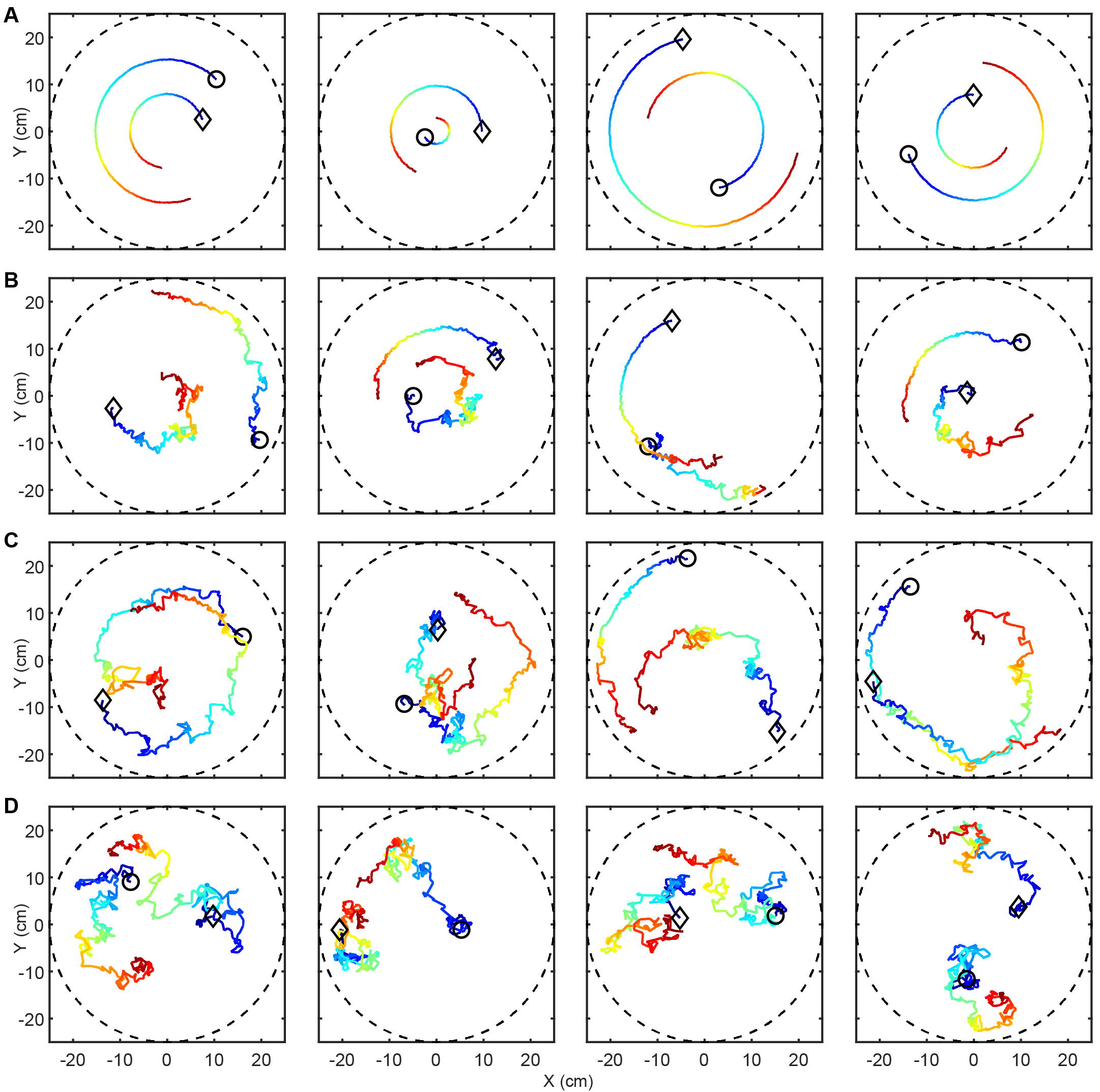} 
    \caption{
\textbf{A-D}: Extended trajectories of particles at $\varphi=0.835, 0.719, 0.699, 0.662$, corresponding to the panels A-D, respectively. Colors from blue to red represent the time from 0 to 120 minutes. The black dashed line indicates the circular boundary of the system, and the diamond and circle markers denote the starting positions of different particles.
}
    \label{SM4}
\end{figure}

In Fig.~\ref{SM4}, we present extended trajectory data over approximately $40$ minutes for several packing fractions, namely $\varphi = 0.835, 0.719, 0.699,$ and $0.662$. These results further corroborate the behavior reported in the main text.

\subsection{Extended analysis of the mean-square displacement}

\begin{figure}
    \centering
    \includegraphics[width=0.85\linewidth]{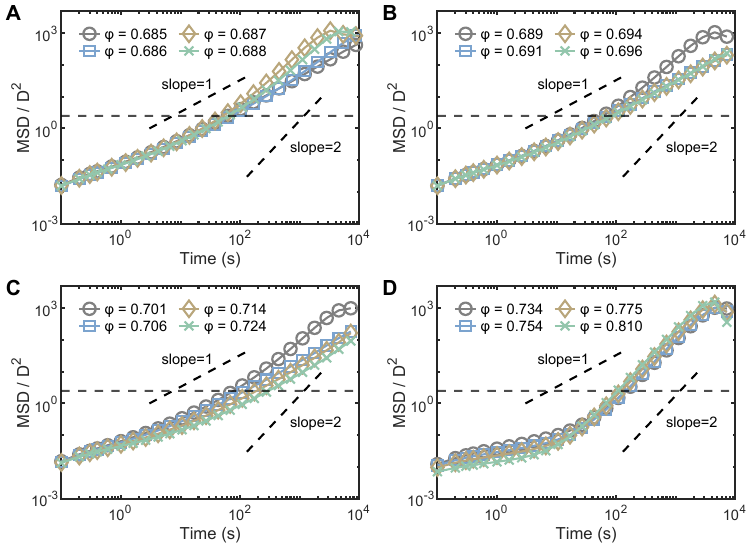} 
    \caption{
\textbf{A-D}: Extended $MSD/D^2$ data for different packing fractions. The horizontal dashed line in each panel represents the crossing value of $MSD/D^2(t^*) = 2.5$, from which the characteristic time $t^*$ is extracted to calculate the $D^2_{min}$. When $\varphi > 0.687$, for $t > t^*$, the results are super-diffusive.
}
    \label{SM12}
\end{figure}

In Fig.~\ref{SM12}, we provide a systematic analysis of the mean-square displacement (MSD) as a function of time across different packing fractions $\varphi$, spanning from the high-density spacetime-crystalline phase to the low-density fluid regime.

Upon increasing the packing fraction, we observe a clear transition from late-time diffusive behavior to a ballistic regime characterized by $\sim t^2$ scaling. This crossover reflects the emergence of global periodic rotation at high packing fractions.

To substantiate this interpretation, we consider a minimal model of global rigid-body rotation. Take a point on a rigid disk undergoing uniform rotation with angular velocity $\Omega$, located at radial distance $r$ from the center. Its trajectory reads
\begin{equation}
\mathbf{r}(t) = r \cos(\Omega t + \Theta)\,\mathbf{e}_x 
+ r \sin(\Omega t + \Theta)\,\mathbf{e}_y,
\label{eq:SM11-1_traj}
\end{equation}
where $\Theta$ is the initial phase.

The displacement relative to the initial position is
\begin{equation}
\mathbf{r}(t) - \mathbf{r}(0)
= r\big(\cos(\Omega t + \Theta) - \cos \Theta\big)\,\mathbf{e}_x
+ r\big(\sin(\Omega t + \Theta) - \sin \Theta\big)\,\mathbf{e}_y.
\label{eq:SM11-2_disp}
\end{equation}
The corresponding squared displacement becomes
\begin{align}
\big|\mathbf{r}(t) - \mathbf{r}(0)\big|^2
&= r^2\big(\cos(\Omega t + \Theta) - \cos \Theta\big)^2
+ r^2\big(\sin(\Omega t + \Theta) - \sin \Theta\big)^2 = 2 r^2 \big(1 - \cos(\Omega t)\big).
\label{eq:SM11-3_dr2}
\end{align}

Assuming particles are uniformly distributed within a disk of radius $R$, the MSD is obtained by averaging Eq.~\eqref{eq:SM11-3_dr2} over the disk area:
\begin{align}
\mathrm{MSD}(t)
&= \frac{1}{\pi R^2} \int_{0}^{R} \int_{0}^{2\pi}
2 r^2 \big(1 - \cos(\Omega t)\big)\, r\, d\theta\, dr = R^2 \big(1 - \cos(\Omega t)\big).
\label{eq:SM11-4_msd}
\end{align}

In the short-time limit ($\Omega t \ll 1$), expanding to second order yields
\begin{equation}
\mathrm{MSD}(t) \approx \frac{R^2}{2}\,\Omega^2 t^2 \propto t^2.
\label{eq:SM11-5_shorttime}
\end{equation}

\begin{figure}
    \centering
    \includegraphics[width=0.85\linewidth]{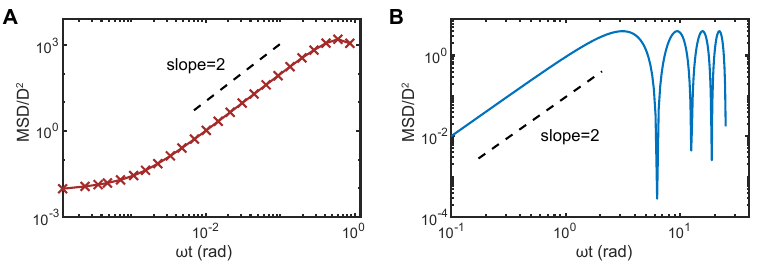} 
    \caption{
   \textbf{A}. Mean-squared displacement (MSD) extracted from the experimental data at packing fraction $\varphi = 0.835$. \textbf{B}. Corresponding MSD computed from an ideal rigid-body rotation model. Here,  $\Omega$ denotes the global angular velocity. The slope equal to $2$ confirms the ballistic ($\sim t^2$) scaling.}
    \label{SM11}
\end{figure}

As summarized in Fig.~\ref{SM11}, this simple rigid-rotation model quantitatively captures the quadratic MSD scaling observed at high packing fractions, confirming that its origin lies in the coherent rigid-body rotation characteristic of the time-crystalline phase.

\subsection{Extended analysis on particle dynamics}
\begin{figure}
    \centering
    \includegraphics[width=0.5\linewidth]{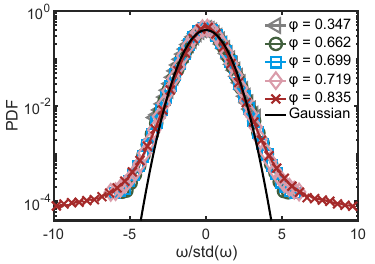} 
    \caption{
Probability function distribution of angular velocities for different packing fractions. The solid black curve represents the standard normal (Gaussian) distribution. The angular velocities are normalized by their respective standard deviations to enable a direct comparison with the standard normal.
}
    \label{SM5}
\end{figure}

In Fig.~\ref{SM5}, we present a statistical analysis of the particle angular velocities for different packing fractions $\varphi$. For all values of $\varphi$, the probability density function (PDF) is symmetric under $\omega \to -\omega$ and centered at $\omega = 0$.

For all packing fractions considered, the angular-velocity distributions exhibit clear deviations from Gaussian behavior in the tails. These non-Gaussian features become increasingly pronounced at higher packing fractions, where progressively heavier tails are observed.

\begin{figure}
    \centering
    \includegraphics[width=0.8\linewidth]{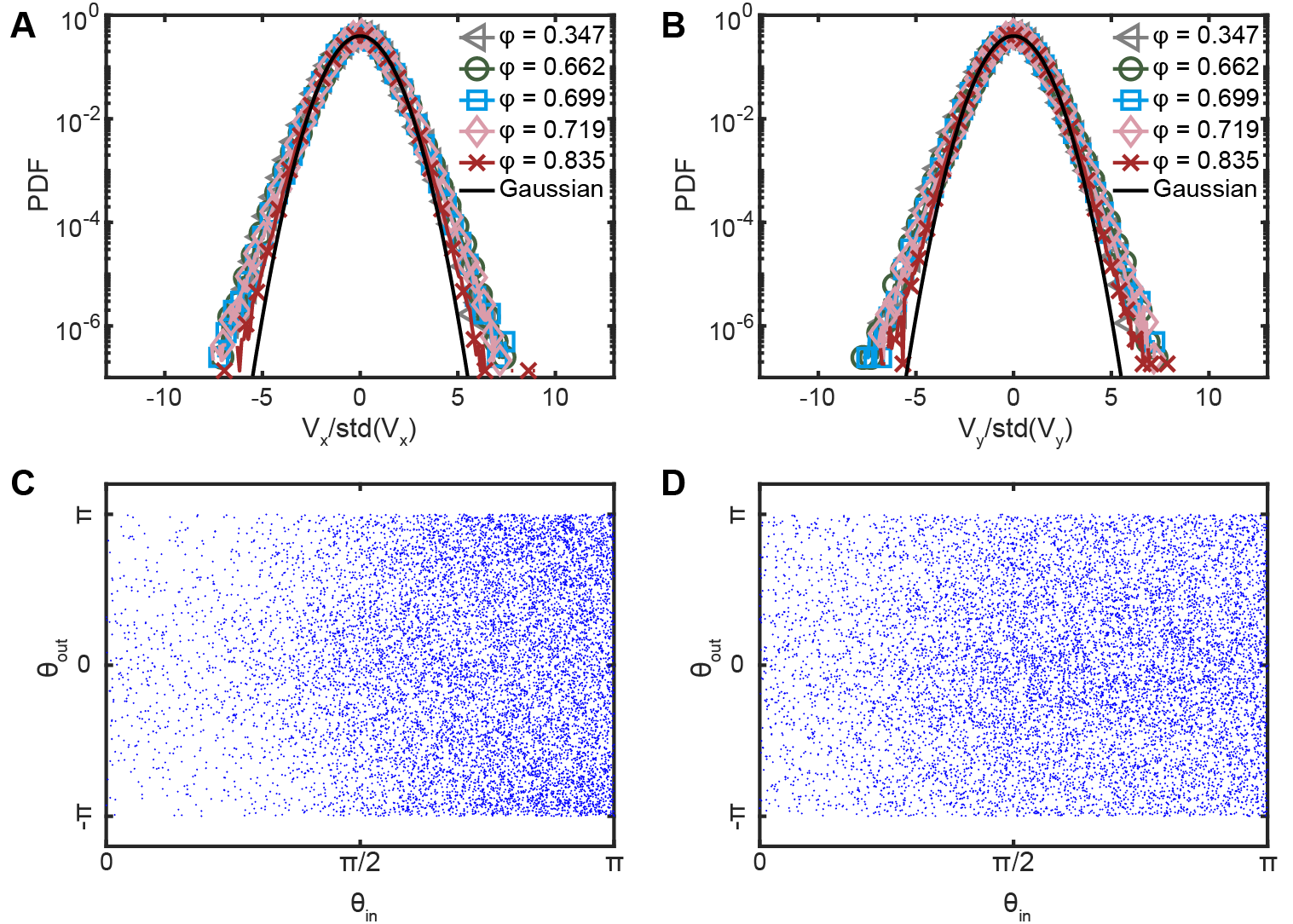} 
    \caption{
   \textbf{A-B}: Distribution of the $x$- and $y$-components of the particle velocities for different packing fractions. The solid black curve represents the standard normal (Gaussian) distribution. \textbf{C-D}: Collision statistics for binary particle encounters. The abscissa ($\theta_{\mathrm{in}}$) denotes the relative angle between the particle velocities immediately before contact, while the ordinate ($\theta_{\mathrm{out}}$) denotes the corresponding relative angle immediately after contact. The pronounced horizontal gradient indicates that collisions occur preferentially for antiparallel incoming velocities ($\theta_{\mathrm{in}} \approx \pi$), whereas nearly parallel velocities ($\theta_{\mathrm{in}} \approx 0$) collide far less frequently. In contrast, no systematic structure is observed along $\theta_{\mathrm{out}}$, indicating the absence of measurable post-collision velocity alignment. Trajectories were sampled at $\Delta t = 0.025\,\mathrm{s}$. A collision event was identified when the interparticle separation changed from larger than tip diameter ($9.7$ mm) immediately before the event to smaller than the pitch diameter ($D=8.8$ mm), at which point the particles were considered to be in contact. Panel \textbf{C} corresponds to $\varphi = 0.347$, and panel \textbf{D} to $\varphi = 0.536$, each based on $10^4$ collision events.}
    \label{SM6}
\end{figure}

In Fig.~\ref{SM6}, we provide additional characterization of the single-particle dynamics, including the PDFs of the $x$- and $y$-components of the velocities and a statistical analysis of binary collision events. The latter demonstrates the absence of any measurable velocity-alignment effect induced by collisions. Further details on the collision analysis are given in the following section.

\subsection{Analysis of particle collisions}
Two additional experiments were conducted in the low-packing fluid phase to examine in detail collision events between pairs of granular disks. The packing fractions were set to $\varphi = 0.347$ and $\varphi = 0.473$. Both experiments yield consistent results, as shown in Fig.~\ref{SM8} and Supplementary Movie~\ref{vid:collision}.

During collisions, the clockwise self-spin angular velocity increases. At the same time, both the translational kinetic energy and the self-spin rotational kinetic energy rise. Collisions transiently constrain the particle–vibrator contact angle, thereby enhancing the efficiency of energy transfer from the external vertical vibration.
\subsection{Velocity auto-correlation function and liquid-like to fluid-like crossover}
Although the dilute, low–packing-fraction regime is not the primary focus of this work, we have carried out a brief analysis of the particle dynamics in this limit. In particular, we report the velocity autocorrelation function (VACF) for several packing fractions, down to the extremely dilute case $\varphi = 0.013$. At the lowest packing fraction, the VACF decays monotonically in time and is well described by a simple exponential form, as expected for a non-interacting or weakly interacting gas.

Interestingly, in the range $0.473 < \varphi < 0.536$, the VACF develops a non-monotonic behavior and exhibits a clear minimum at approximately $t \approx 0.07\,\mathrm{s}$. According to Frenkel's criterion~\cite{PhysRevLett.111.145901}, this qualitative change signals the dynamical crossover from a gas-like to a liquid-like regime, consistent with observations in related systems~\cite{jiang2025experimental}.

\subsection{Extended analysis on structural probes}
In Fig.~\ref{SM9}, we present extended data for the sixfold bond-orientational correlation function $G_6(r)$. Panel A reproduces the behavior across the four distinct phases discussed in the main text.

In the crystalline phase ($\varphi > 0.734$), orientational correlations are long-ranged and do not decay with distance. In the T-coexistence region ($0.709 < \varphi < 0.734$), the system exhibits hexatic order, and $G_6(r)$ follows the expected power-law decay. For smaller packing fractions, correlations become short-ranged and display a clear exponential decay.

The corresponding correlation length $\xi_6$ is extracted by fitting the long-distance behavior to the exponential form $G_6(r) \sim \exp(-r/\xi_6)$. The resulting dependence of $\xi_6$ on packing fraction is shown in  Fig.~\ref{SM9}B. The correlation length decreases systematically as $\varphi$ is reduced. Around $\varphi_1 \approx 0.687$, where hexatic order is completely lost and the system is fully fluid, the orientational correlation length $\xi_6$ becomes microscopic, of the order of the particle diameter $D$.

In Fig.~\ref{SM13}A, we show the translational correlation function $G_T(r)$ for the same four benchmark packing fractions. At high packing, $G_T(r)$ exhibits a slow power-law decay, as expected for a two-dimensional crystalline phase. In the hexatic regime, the power-law exponent drops below $-1/3$, in agreement with KTHNY theory \cite{2dmelt-RevModPhys.60.161}, and eventually crosses over to a short-range exponential decay as the system enters the fluid phase.

For completeness, panel B of the same figure displays the radial distribution function $g(r)$, which likewise captures the transition from an ordered crystalline state to a disordered fluid phase.

\subsection{Topological defect analysis}
Delaunay triangulation is employed to construct the neighbor network from the 2D particle positions. A particle is classified as defective if its coordination number differs from six. We refer to the Methods for a more detailed discussion about this classification and to Fig.~\ref{fig:defect} for a visual representation of the most relevant defects.

To characterize spatial organization, the system is partitioned into a grid of boxes covering the entire domain, with each box having a linear size approximately $1.5$ times the mean interparticle spacing. A box is labeled as defective if it contains at least one defective particle. Two boxes are considered connected if they share either an edge or a corner, and a connected cluster is defined as a set of defective boxes linked through such connections.

For each cluster $i$, consisting of $n_c^{(i)}$ defective boxes located at positions ${\mathbf{r}_{1},\mathbf{r}_{2},\dots,\mathbf{r}_{{n_c}^{(i)}}}$, the center of mass is defined as
\begin{equation}\label{Topological_defect_analysis_1}
\mathbf{r}_\text{cm}^{(i)}=\frac{1}{n_c^{(i)}}\sum_{j=1}^{n_c^{(i)}}\mathbf{r}_{j},
\end{equation}
and the corresponding radius of gyration is
\begin{equation}\label{Topological_defect_analysis_2}
R_{g}^{(i)}=\sqrt{\frac{1}{n_{c}^{(i)}}\sum_{j=1}^{n_{c}^{(i)}}\left|\mathbf{r}_{j}-\mathbf{r}_\text{cm}^{(i)}\right|^{2}}.
\end{equation}

The data pairs $(\log_{10} R_g, \log_{10} n_c)$ are collected from approximately 1000 randomly selected snapshots of particle configurations. Clusters containing three or fewer defective boxes are excluded from the analysis. The fractal dimension $d_f$ is extracted from the slope of a linear regression performed over all data points:
\begin{equation}
\log_{10} n_c = d_f,\log_{10} R_g + C,
\label{eq: fractal}
\end{equation}
where $C$ is a fitting constant and $d_f$ corresponds to the fractal dimension.

Physically, $d_f=0$ indicates point-like structures, $d_f=1$ linear (string-like) clusters, and $d_f=2$ area-filling clusters. A cluster is identified as percolating if it contains antipodal defective boxes that simultaneously touch the system boundary.

In Fig.~\ref{SM10} we provide further details of this analysis.

In panel A, we show a direct visual correlation between the location of defective particles and regions where the six-fold bond orientational order parameter $|\Psi_6|$ is low. Panel B illustrates the procedure used to identify and define defect clusters. In panel C, we present a representative configuration in which free disclinations, free dislocations, dislocation pairs (quadrupoles), and defect clusters are indicated by different colored symbols. As clearly visible, the system exhibits several aggregated clusters of defects.

In panel D, we show the scaling of the number of clusters $n_c$ as a function of their gyration radius $R_{gc}$, which displays an approximately linear behavior in the regime considered. Panel E reports the corresponding fractal dimension $d_f$ of the defect clusters as a function of the packing fraction. In the crystalline regime, we find $d_f \approx 0.5$, indicating an ultra-dilute population of defects. This value is consistent with the presence of a small number of sparse and essentially uncorrelated defect points. Upon decreasing the packing fraction and entering the coexistence regions, the fractal dimension increases to $1 < d_f < 2$, signaling the formation of connected defect clusters whose effective dimensionality grows as $\varpi$ decreases. Near the transition to the fluid phase, $d_f$ approaches the two-dimensional limit $d_f = 2$, consistent with the emergence of extended, space-filling defect clusters.

Finally, panel F shows the percolation ratio of these clusters. This ratio remains close to zero throughout the crystalline regime and only begins to increase when fluid regions emerge in the low-packing coexistence region. It then rises rapidly and approaches unity in the fully developed fluid phase, indicating complete percolation of defect clusters.

\subsection{Goldstone mode analysis} 
\textbf{Definition of Phase Fluctuations $\phi(t)$} \\
The phase fluctuation $\phi_i(t)$ for a particle labeled $i$ represents the angular deviation of that particle's phase angle from its global rotation. Positions are defined relative to the system center $(x_c, y_c)$. The instantaneous raw phase angle for a particle labeled as $i$ at a certain time step $t$, $\theta_{raw, i}(t)$, is calculated as:
\begin{equation}\label{Goldstone_mode_analysis_1}
    \theta_{raw, i}(t) = \operatorname{arctan}\left(y_i(t) - y_c, \; x_i(t) - x_c\right)
\end{equation}

To track total rotation, the raw angle is unwrapped to remove $2\pi$ discontinuities:
\begin{equation}\label{Goldstone_mode_analysis_2}
    \theta_{total, i}(t) = \operatorname{Unwrap}(\theta_{raw, i}(t))
\end{equation}

Then, the mean rotation is removed by fitting a linear trend to $\theta_{total, i}(t)$. The average angular velocity $\bar{\omega}_i$ is fitted individually for each particle,
\begin{equation}\label{Goldstone_mode_analysis_3}
    \theta_{fit, i}(t) = \bar{\omega}_i t + \theta_{0, i}
\end{equation}
and the corresponding phase fluctuation $\phi_i(t)$ for each particle is the residual of this fit:
\begin{equation}\label{Goldstone_mode_analysis_4}
    \phi_i(t) = \theta_{total, i}(t) - \theta_{fit, i}(t)
\end{equation}

\noindent \textbf{The calculation of static structure factor $S_\phi(\mathbf{q})$}

The static structure factor $S_\phi(\mathbf{q})$ for phase fluctuations is calculated via
\begin{equation}\label{Goldstone_mode_analysis_5}
S_\phi(\mathbf{q}) = \frac{1}{N} \left\langle \left| \sum_{j=1}^{N} \phi_j(t) e^{-i \mathbf{q} \cdot \mathbf{r}_j} \right|^2 \right\rangle_t
\end{equation}
Here $\mathbf{q}$ is the wave vector, $\phi_j(t)$ represents the phase fluctuations at time $t$ for a particle labeled by $j$ by detrending the global rotation, as shown in Fig. \ref{S91}, and $\left\langle \right\rangle_t$ means ensemble average over time.

Due to its gapless nature, the Goldstone mode would induce an infrared divergence in the phase static structure factor $S_{\phi}(q)\sim q^{-\alpha}$ where $\alpha$ depends on the dimensions of the system and the dispersion relation of the corresponding Goldstone mode. In Fig.~\ref{S92}, we provide the experimental results for the phase static structure factor at different packing fractions. In the spacetime crystal phase, we observe a divergent behavior of this quantity in the limit of zero wavevector that is consistent with a $1/q^2$ scaling. This observation indicates the emergence of a gapless mode, which we identify as the Goldstone mode for time translations.

\noindent\textbf{The calculation of dynamic structure factor $S_\phi(\mathbf{q}, \omega)$}
In order to corroborate more firmly the emergence of a gapless Goldstone mode, we move to the dynamic structure factor for the phase fluctuations.

The dynamic structure factor $S_\phi(\mathbf{q}, \omega)$ is calculated by first constructing the correlation matrix for phase fluctuations $C_{ij}^{\phi}$, where
\begin{equation}\label{Goldstone_mode_analysis_6}
    C^{\phi}_{ij} = \langle \phi_i(t) \phi_j(t) \rangle_t
\end{equation}
for the phase $\phi$ of particles in the system. The dynamical matrix $D_M$ is then the inverse matrix of $C_{ij}^{\phi}$. An eigenvalue decomposition is performed to find the "normal modes" of $D_M$, 
\begin{equation}\label{Goldstone_mode_analysis_7}
    D_M \mathbf{v}_n = \lambda_n \mathbf{v}_n
\end{equation}
The frequency of the mode is related to the eigenvalues $\lambda_n$ via $\omega_n=\sqrt{\lambda_n}$. The eigenvector is Fourier transformed into components with wavevector $\mathbf{q}$, namely, $A_{n}(\mathbf{q}) = \sum_{j=1}^{N} v_{n, j} e^{-i \mathbf{q} \cdot \mathbf{r}_{j}}$. The dynamical structure factor $S_\phi(\mathbf{q}, \omega)$ is then the the summation of contributions of all modes at frequency $\omega$
\begin{equation}\label{Goldstone_mode_analysis_8}
S_\phi(\mathbf{q}, \omega) = \sum_{n} \delta(\omega - \omega_{n}) \cdot |A_{n}(\mathbf{q})|^{2}
\end{equation}

Because the covariance matrix $C^{\phi}_{ij}$ is constructed exclusively
from equal-time spatial correlations of the phase $\phi$, the resulting mode frequencies
$\omega_n$ are strictly dimensionless and do not carry an intrinsic time scale.
To express them in physical units, they must be rescaled by a characteristic
kinetic rate of the system.

The instantaneous phase velocity of each particle is computed as the
discrete time derivative of its phase fluctuation:
\begin{equation}
\label{Goldstone_mode_analysis_9}
v_{\phi,i}(t)
=
\frac{\phi_i(t+\Delta t)-\phi_i(t)}{\Delta t},
\end{equation}
where $\Delta t$ is the time interval between consecutive frames.

We define a characteristic kinetic rate $\Gamma$ as the ensemble and
time average of the absolute phase velocities:
\begin{equation}
\label{Goldstone_mode_analysis_10}
\Gamma
=
\left\langle
|v_{\phi,i}(t)|
\right\rangle_{i,t}.
\end{equation}

The dimensionless pseudo-frequencies $\omega_n$ are then converted
into physical angular frequencies (in units of $\text{s}^{-1}$) through
\begin{equation}
\label{Goldstone_mode_analysis_11}
\omega^{\mathrm{phys}}_n
=
\Gamma \,\omega_n.
\end{equation}

This rescaling restores the appropriate temporal scale and enables
direct comparison with experimentally measured dynamical spectra.

The results of $S_\phi(\mathbf{q}, \omega)$ for various packing fraction values $\varphi$ are shown in Fig. \ref{S93}. Once again, at high density, in the spacetime crystal phase, we observe the emergence of a gapless mode with approximate dispersion:
\begin{equation}\label{Goldstone_mode_analysis_9}
    \omega= v_\phi q -i \gamma(q)+\dots,
\end{equation}
where the exact form of the damping $\gamma$ is difficult to determine from the current data as shown in Fig. \ref{S93}, which consistently approaches zero as $q$ approaches zero, as expected for a Goldstone mode.
This indicates the emergence of an underdamped propagating gapless mode that is consistent with the Goldstone degree of freedom. Interestingly, we observe that its velocity $v_\phi$ increases with increasing $\varphi$, a manifestation of the increasing stiffness of the spacetime crystalline state at higher packing fractions. 

Finally, we remark that the propagating and linearly dispersing nature of this mode is the result of the concomitant breaking of time and space translations.

\subsection{Caption for Movies}

\newcounter{suppvideo}
\renewcommand{\thesuppvideo}{M\arabic{suppvideo}}

\noindent\refstepcounter{suppvideo}\label{vid:835}
\textbf{Movie \thesuppvideo.}
(\href{run:SM_Movie1.mp4}{Movie \thesuppvideo}).
Experiment with RS particles, as introduced in SM3, at $\varphi = 0.835$. The trajectories of three representative particles are shown over 40 minutes, exhibiting regular collective motion characteristic of a time-crystal state.

\noindent\refstepcounter{suppvideo}\label{vid:MRS}
\textbf{Movie \thesuppvideo.}
(\href{run:SM_Movie2.mp4}{Movie \thesuppvideo}).
Experiment with MRS particles, as introduced in SM3, at $\varphi = 0.835$. The global rotation direction is counterclockwise.

\noindent\refstepcounter{suppvideo}\label{vid:CS}
\textbf{Movie \thesuppvideo.}
(\href{run:SM_Movie3.mp4}{Movie \thesuppvideo}).
Experiment with CS particles, as introduced in SM3, at $\varphi = 0.835$. The global rotation direction is counterclockwise.

\noindent\refstepcounter{suppvideo}\label{vid:RC}
\textbf{Movie \thesuppvideo.}
(\href{run:SM_Movie4.mp4}{Movie \thesuppvideo}).
Experiment with RC particles, as introduced in SM3, at $\varphi = 0.835$. The global rotation direction is clockwise.

\noindent\refstepcounter{suppvideo}\label{vid:noise}
\textbf{Movie \thesuppvideo.}
(\href{run:SM_Movie5.mp4}{Movie \thesuppvideo}).
Experiment with RS particles (see SM3) at $\varphi = 0.806$. A loudspeaker was placed between the vibrator and the sample to introduce external acoustic noise (see SM2). The speaker operated at maximum power; nevertheless, the time-crystal behavior persists.

\noindent\refstepcounter{suppvideo}\label{vid:random_phase}
\textbf{Movie \thesuppvideo.}
(\href{run:SM_Movie6.mp4}{Movie \thesuppvideo}).
Seven small systems, each 15\,cm in diameter, demonstrate spontaneous breaking of time-translation symmetry and the emergence of random phases. See SM1 for details.

\noindent\refstepcounter{suppvideo}\label{vid:662}
\textbf{Movie \thesuppvideo.}
(\href{run:SM_Movie7.mp4}{Movie \thesuppvideo}).
Experiment with RS particles at $\varphi = 0.662$. The trajectories of three representative particles are shown over 40 minutes. The system exhibits neither radial nor azimuthal order and has transitioned into a disordered state.

\noindent\refstepcounter{suppvideo}\label{vid:699}
\textbf{Movie \thesuppvideo.}
(\href{run:SM_Movie8.mp4}{Movie \thesuppvideo}).
Experiment with RS particles at $\varphi = 0.699$. The trajectories of three representative particles are shown over 40 minutes. Radial ordering is largely lost, with only azimuthal order remaining, while counterclockwise global rotation persists.

\noindent\refstepcounter{suppvideo}\label{vid:719}
\textbf{Movie \thesuppvideo.}
(\href{run:SM_Movie9.mp4}{Movie \thesuppvideo}).
Experiment with RS particles at $\varphi = 0.719$. The trajectories of three representative particles are shown over 40 minutes. The upper-left region displays regular collective motion, while the lower-right region shows a partial loss of radial ordering.

\begin{figure}
    \centering
    \includegraphics[width=\linewidth]{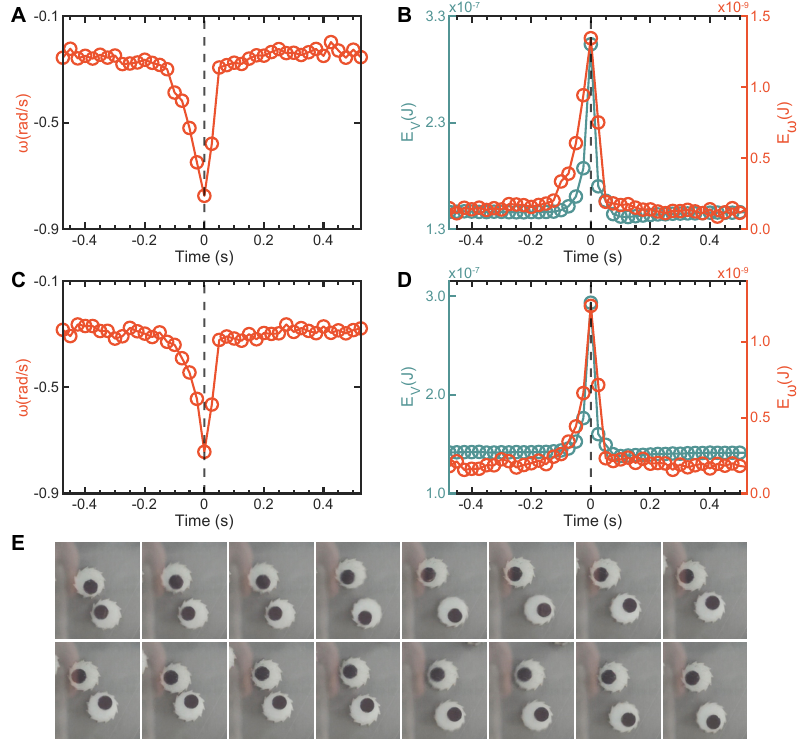} 
    \caption{
    \textbf{A}. Particle-averaged self-spin angular velocity during a collision event at $\varphi = 0.347$. \textbf{B}. Particle-averaged translational kinetic energy and self-spin rotational kinetic energy during the same collision at $\varphi = 0.347$. \textbf{C}. Particle-averaged self-spin angular velocity during a collision event at $\varphi = 0.473$. \textbf{D}. Particle-averaged translational kinetic energy and self-spin rotational kinetic energy during the same collision at $\varphi = 0.473$. \textbf{E}. Snapshot of a representative collision event. Prior to impact, the lower particle rotates counterclockwise; after the collision, its rotation reverses to clockwise. The video was recorded with a mobile phone at $30$ fps and it is provided as Supplementary Movie \hyperref[vid:collision]{~\ref*{vid:collision}}. Negative values of $\omega$ correspond to clockwise self-spin rotation. }
    \label{SM8}
\end{figure}

\noindent\refstepcounter{suppvideo}\label{vid:collision}
\textbf{Movie \thesuppvideo.}
(\href{run:SM_Movie10.mp4}{Movie \thesuppvideo}).
Video corresponding to the collision event shown in Fig.~\ref{SM8}. The bottom-right particle rotates counterclockwise before impact and reverses to clockwise rotation afterward.

\newpage
\clearpage


\begin{figure}
    \centering
    \includegraphics[width=0.85\linewidth]{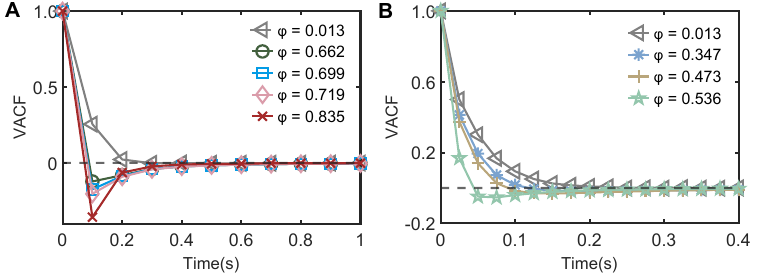} 
    \caption{\textbf{A–B} Normalized velocity autocorrelation functions (VACF). The acquisition frame rates are $10$ fps in panel \textbf{A} and $40$ fps in panel \textbf{B}. For $0.473 < \varphi < 0.536$, a pronounced minimum emerges around $t \approx 0.7\,\mathrm{s}$. According to the Frenkel criterion~\cite{PhysRevLett.111.145901}, this feature signals the dynamical crossover from a fluid-like to a liquid-like regime.}
    \label{SM7}
\end{figure}

\begin{figure}
    \centering
    \includegraphics[width=0.85\linewidth]{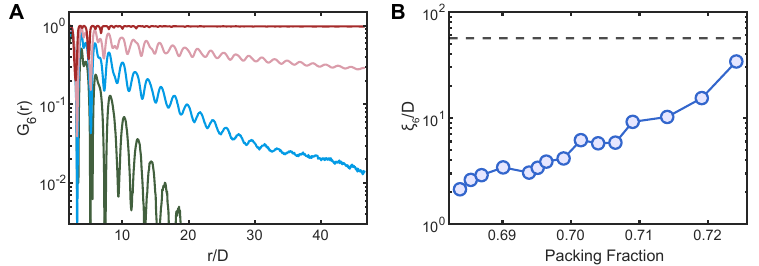} 
    \caption{
   \textbf{A}. Spatial hexatic correlation function $G_6(r)$. Four representative curves are shown, each corresponding to a distinct phase and selected according to the same color scheme used in the main text. \textbf{B}. Correlation length $\xi_6$ extracted from $G_6(r)$ by fitting the long-distance decay to an exponential form, $G_6(r) \sim \exp(-r/\xi_6)$, where $\xi_6$ is defined as the characteristic decay length. The horizontal dashed line indicates the system size, $L = 56.8D$.
}
    \label{SM9}
\end{figure}
\begin{figure}
    \centering
    \includegraphics[width=0.85\linewidth]{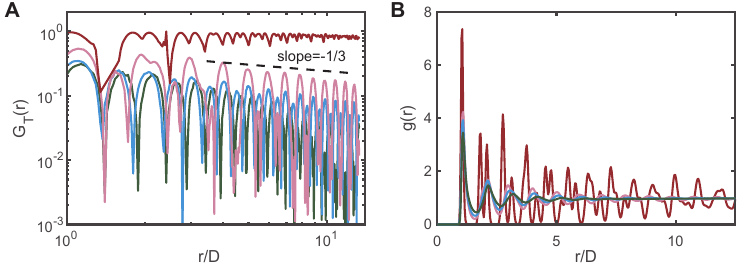} 
    \caption{
   \textbf{A}. Translational correlation function $G_T(r)$ for four selected packing fractions $\varphi$ corresponding to the four phases described in the main text. The dashed line indicates a slope of $-1/3$, corresponding to the KTHNY prediction at the solid-hexatic transition. \textbf{B}. Radial distribution function $g(r)$ for the same four values of $\varphi$. In the spacetime-crystal phase (brick red), $g(r)$ exhibits pronounced triangular-lattice peaks at $D$, $\sqrt{3}D$, $2D$, and $\sqrt{7}D$. Upon melting, these crystalline signatures disappear, and $g(r)$ rapidly approaches unity.}
    \label{SM13}
\end{figure}

\begin{figure}
    \centering
    \includegraphics{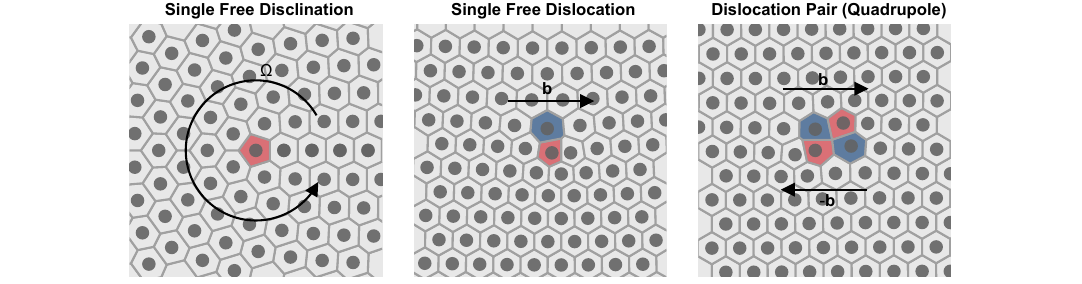}
    \caption{Classification of topological defects in a triangular lattice. (a) Isolated disclination ($5$-fold in red). (b) A dislocation (bound dipole) with a non-zero Burgers vector $\mathbf{b}$. (c) A dislocation pair (quadrupole) localizing strain with canceling Burgers vectors.}
    \label{fig:defect}
\end{figure}

\begin{figure}
    \centering
    \includegraphics[width=\textwidth]{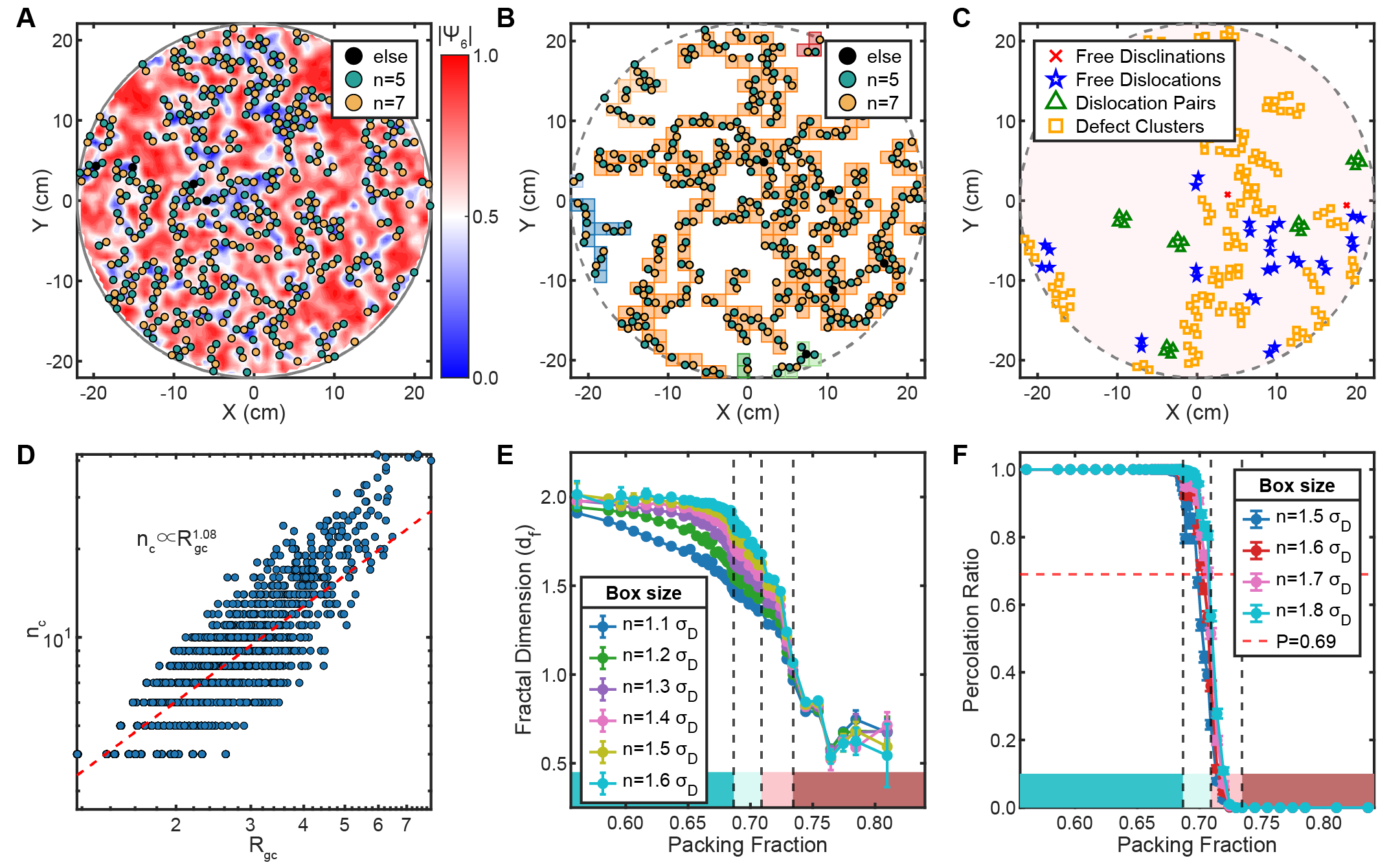} 
    \caption{
    \textbf{A}: {Correlation between the spatial distribution of defects and $|\Psi_6|$. Red regions correspond to high values of $|\Psi_6|$, while blue regions indicate low $|\Psi_6|$. Via Voronoi tessellation, defective particles are colored according to their neighbor counts: green for particles with 5 neighbors, orange for those with 7 neighbors, and black for particles with any number of neighbors other than 6. Defective particles are predominantly located in regions of low $|\Psi_6|$. \textbf{B}: Identification of defect clusters. A box is labeled as defective if it contains at least one defective particle. A defect cluster is defined as a group of boxes that are edge- or diagonally adjacent. A cluster is classified as percolating if it contains defective boxes that touch opposite boundaries of the system. The cluster highlighted in orange is therefore percolating. \textbf{C}: Identification of dislocations, disclinations, and defect clusters. Free dislocations and free disclinations denote isolated topological defects surrounded by particles with regular coordination. A dislocation pair (quadrupole) consists of two bound dislocations. Dislocations are marked by pairs of blue stars, dislocation pairs by four green triangles, and free disclinations by crosses. \textbf{D}: Linear regression used to extract the fractal dimension, following Eq.~\eqref{eq: fractal}. Each data point represents the cluster size (i.e., the number of defective boxes) plotted against the corresponding radius of gyration. The data are collected from 1000 randomly selected frames at $\varphi = 73.4\%$. \textbf{E}: Fractal dimension as a function of $\varphi$ for different box sizes. The fractal dimension approaches $d_f = 2$ on the fluid side and is approximately $d_f = 1$ at $\varphi = 73.4\%$. Error bars correspond to the standard deviation of $d_f$ computed over independent data sets. $\sigma_D$ refers to the average inter-particle distance and n refers to the side length of a box. \textbf{F}: Percolation ratio as a function of $\varphi$, computed for different box sizes. The percolation ratio is defined as the fraction of frames (out of 1000 randomly selected configurations at each $\varphi$) that contain a percolating defect cluster. It remains close to zero in the spacetime crystal phase and approaches unity in the fluid phase, with a sharp increase in the coexistence region.}
}
    \label{SM10}
\end{figure}

\begin{figure}
    \centering
    \includegraphics{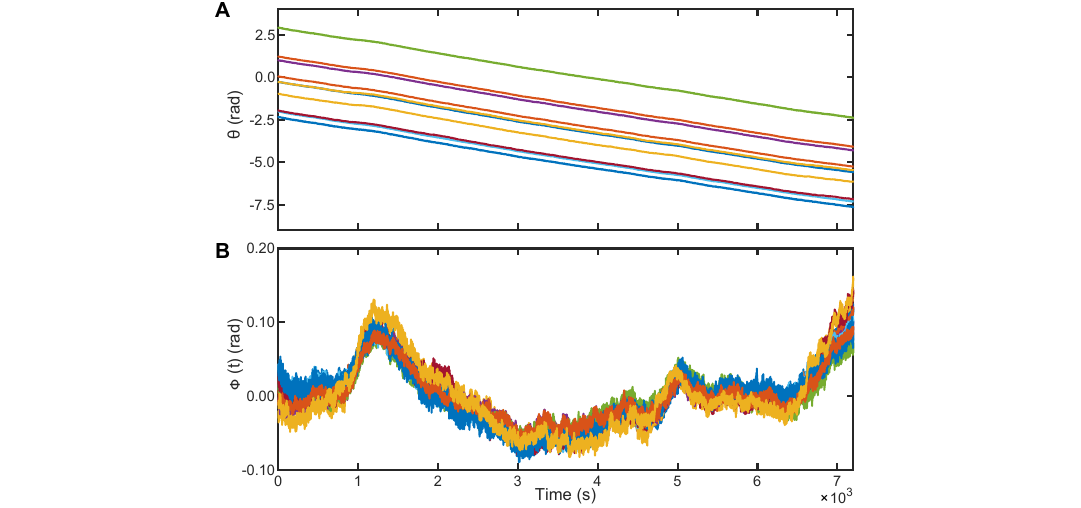} 
    \caption{
   The upper panel shows the original particle phase trajectories, randomly selected at $\varphi = 80.98\%$. The lower panel displays the corresponding phase fluctuations after subtracting the global rotational trend.
}
    \label{S91}
\end{figure}

\begin{figure}
    \centering
    \includegraphics[width=\textwidth]{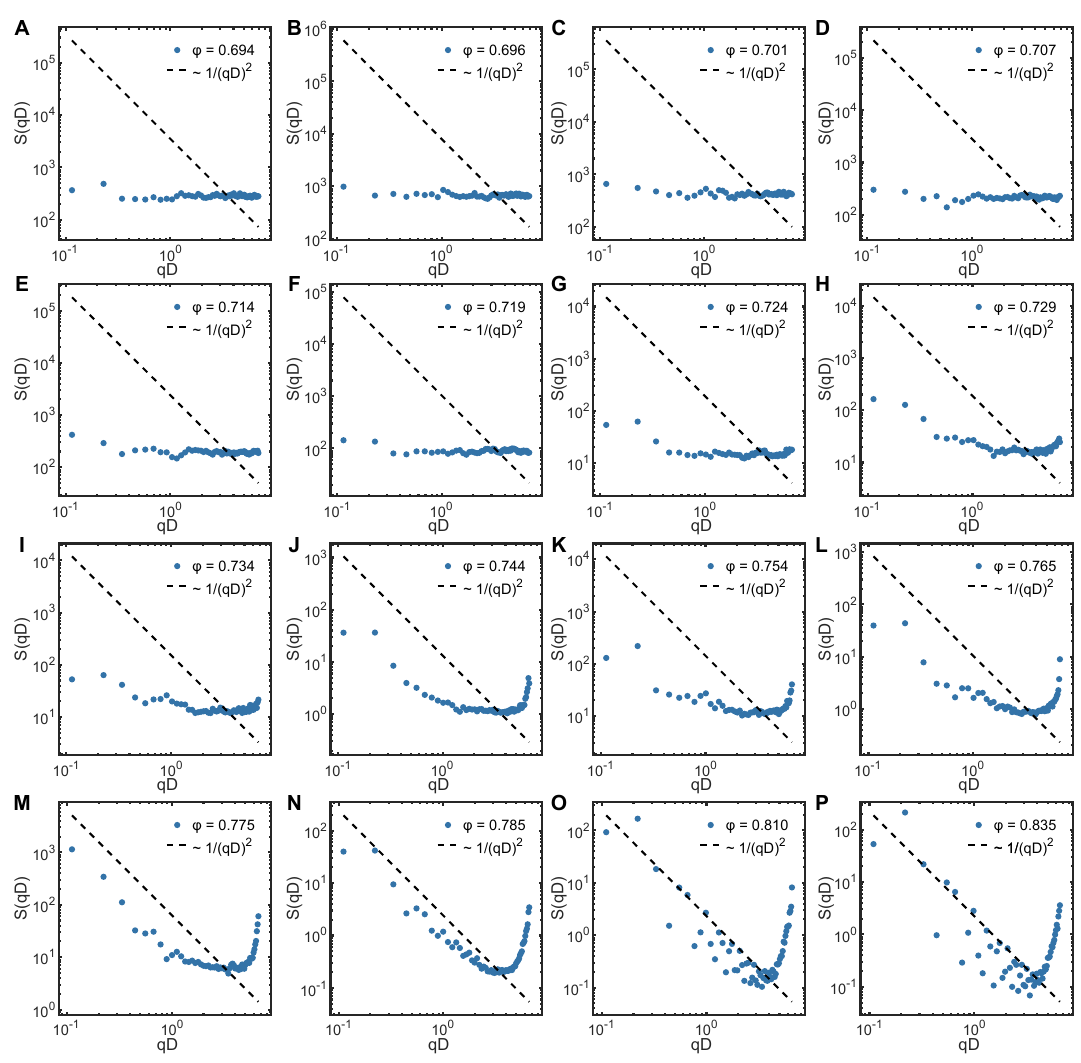} 
    \caption{
    The static structure factor $S(qD)$, where $q = |\mathbf{q}|$ and $D$ denotes the average spacing among particles, is shown for selected values of $\varphi$. The off-diagonal black dashed line indicates the expected Goldstone mode, characterized by a $(qD)^{-2}$ scaling as the wave vector $qD \to 0$. From the data, we infer that the Goldstone mode emerges at the onset of the spacetime crystal phase.}
    \label{S92}
\end{figure}

\begin{figure}
    \centering
    \includegraphics[width=\textwidth]{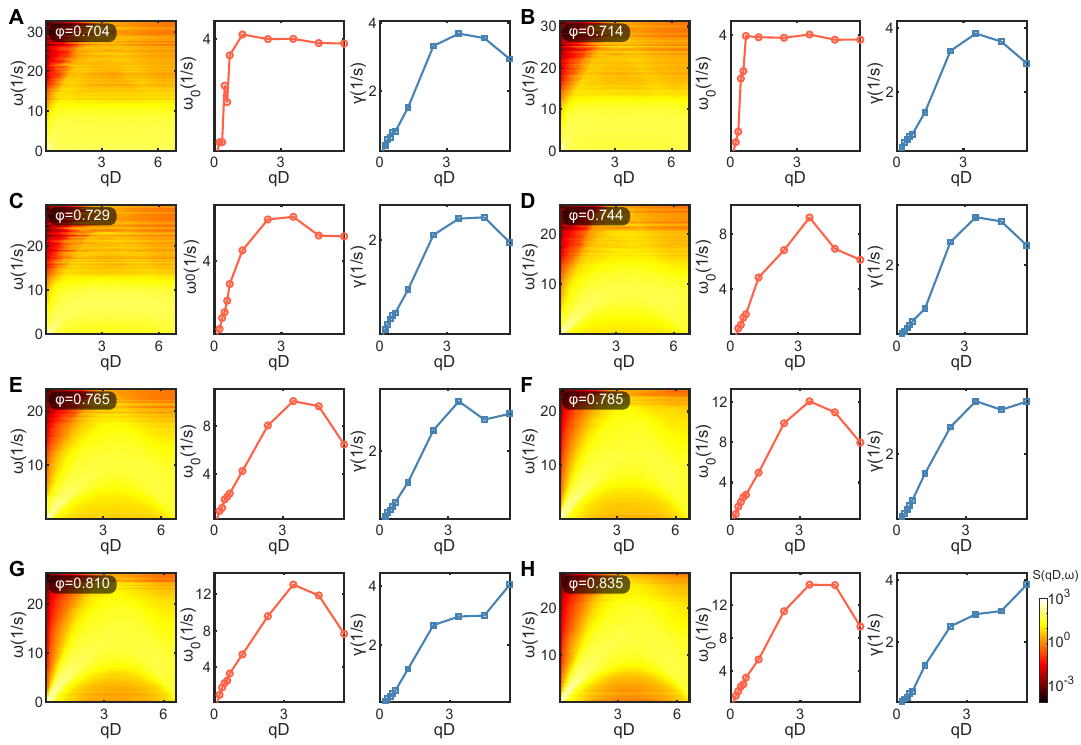} 
    \caption{
    The dynamical structure factor $S(qD,\omega)$, where $q = |\mathbf{q}|$ and $D$ denotes the average spacing among particles, is shown for selected values of $\varphi$. For each packing fraction, the left panel displays the full two-dimensional spectrum. The central and right panels report the extracted central frequency $\omega_0(qD)$ and damping width $\gamma(qD)$, respectively, obtained from a Lorentzian fit of the form $S(qD,\omega)=\frac{A}{\pi}\frac{\gamma}{(\omega-\omega_0)^2+\gamma^2}$ at fixed $qD$. At high packing fractions, the low-$q$ behavior indicates a propagating mode with linear dispersion $\omega_0 \sim q$. The damping $\gamma$ goes consistently to zero as $q$ approaches zero as expected for a Goldstone mode but the exact power is difficult to be obtained from the current data. 
}
    \label{S93}
\end{figure}

\end{document}